\documentclass{aa}
\usepackage{txfonts,textcomp}
\usepackage{graphicx}
\usepackage{color}
\usepackage{tikz}
\usepackage{amssymb}
\usetikzlibrary{shapes,arrows}
\usepackage{multirow}
\usepackage{natbib}
\usepackage{pdflscape}
\bibpunct{(}{)}{;}{a}{}{,}

\begin{document}
\title{The interior rotation of a sample of $\gamma$ Doradus stars from
  ensemble~modelling of their gravity mode period spacings\thanks{Based on data
    gathered with the NASA Discovery mission \emph{Kepler} and the HERMES
    spectrograph, which is installed at the Mercator Telescope, operated on the
    island of La Palma by the Flemish Community at the Spanish Observatorio del
    Roque de los Muchachos of the Instituto de Astrof\'isica de Canarias, and
    supported by the Fund for Scientific Research of Flanders (FWO), Belgium,
    the Research Council of KU Leuven, Belgium, the Fonds National de la
    Recherche Scientifique (F.R.S.-FNRS), Belgium, the Royal Observatory of
    Belgium, the Observatoire de Gen\`eve, Switzerland, and the Th\"uringer
    Landessternwarte Tautenburg, Germany.}}

\author{T.~Van~Reeth\inst{1}
\and A.~Tkachenko\inst{1} 
\and C.~Aerts\inst{1,2}
} 
\institute{Instituut voor Sterrenkunde, KU Leuven, Celestijnenlaan 200D, 3001 Leuven, Belgium
\and Department of Astrophysics, IMAPP, Radboud University Nijmegen, PO Box 9010, 6500 GL Nijmegen, The Netherlands}
\authorrunning{T. Van Reeth et al.}
\titlerunning{The rotation of $\gamma$ Dor stars}

\date{Received / Accepted}
\abstract{{CONTEXT} Gamma Doradus stars (hereafter $\gamma$\,Dor stars) are
  known to exhibit
  gravity- and/or gravito-intertial modes that probe the inner stellar region
  near the convective core boundary. The non-equidistant spacing of the
  pulsation periods is an observational signature of the stars' evolution and
  current internal structure and is heavily influenced by rotation.\\
{AIMS} We aim to constrain the near-core rotation rates for a sample of $\gamma$
Dor stars, for which we have detected period spacing patterns.\\
{METHODS} We combined the asymptotic period spacing with the traditional
approximation of stellar pulsation to fit the observed period spacing patterns
using $\chi^2$-optimisation. The method was applied to the observed period
spacing patterns of a sample of stars and used for ensemble modelling.\\ 
{RESULTS} For the majority of stars with an observed period spacing pattern we
successfully determined the rotation rates and the asymptotic period spacing
values, though the uncertainty margins on the latter were typically large. This
also resulted directly in the identification of the modes corresponding with the
detected pulsation frequencies, which for most stars were prograde dipole gravity
and gravito-inertial modes. The majority of the observed retrograde modes were found to 
be Rossby modes.
We further discuss the limitations of the method due to the neglect of
the centrifugal force and the incomplete treatment of the Coriolis
  force.\\
{CONCLUSION} Despite its current limitations, the proposed methodology was
successful to derive the rotation rates and to identify the modes from the
  observed period spacing patterns. 
It forms the first step towards detailed seismic 
modelling based on observed period spacing patterns of moderately to rapidly 
rotating $\gamma\,$Dor stars.}


\maketitle

\section{Introduction}
\label{sec:intro}
Gamma Dor stars are early F- to late A-type stars (with $1.4\,M_\odot \lesssim M
\lesssim 2.0\,M_\odot$) which exhibit non-radial gravity and/or gravito-inertial
mode pulsations \citep[e.g.][]{Kaye1999}. This places them directly within the
transition region between low-mass stars with a convective envelope and
intermediate-mass stars with a convective core, where the CNO-cycle becomes
increasingly important relative to the pp-chain as the dominant hydrogen burning
mechanism \citep[e.g.][]{Silva2011}. The pulsations in $\gamma$\,Dor stars are
excited by the flux blocking mechanism at the bottom of the convective envelope
\citep{Guzik2000,Dupret2005}, though the $\kappa$ mechanism has been linked to
$\gamma$\,Dor type pulsations as well \citep{Xiong2016}. The oscillations
predominantly trace the radiative region near the convective core boundary. As a
result these pulsators are ideal to characterise the structure of the deep
stellar interior.

As shown by \citet{Tassoul1980}, high order ($n \gg l$) gravity modes are
asymptotically equidistant in period for non-rotating chemically homogeneous
stars with a convective core and a radiative envelope. This study was further
expanded upon by \citet{Miglio2008a}. The authors found characteristic dips to
be present in the period spacing series when the influence of a chemical
gradient is included in the analysis. The periodicity of the deviations is
related to the location of the chemical gradient, while the amplitude of the
dips was found to be indicative of the steepness of the
gradient. \citet{Bouabid2013} further improved upon the study by including the
effects of both diffusive mixing and rotation, which they introduced using the
traditional approximation. The authors concluded that the mixing processes
partially wash out the chemical gradients inside the star, resulting in a
reduced amplitude for the dips in the spacing pattern. Stellar rotation
introduces a shift in the pulsation frequencies, leading to a slope in the
period spacing pattern. Zonal and prograde modes, as seen by an
  observer in an inertial frame of reference, were found to have a downward slope, while the pattern for the
retrograde {high order} modes has an upward slope.

Over the past decade the observational study of pulsating stars has benefitted
tremendously from several space-based photometric missions, such as MOST
\citep{Walker2003}, CoRoT \citep{Auvergne2009} and \emph{Kepler}
\citep{Koch2010}. While typically only a handful of modes could be resolved
using ground-based data, the space missions have provided us with
near-continuous high S/N observations of thousands of stars on a long time base,
resulting in the accurate determination of dozens to hundreds of pulsation
frequencies for many targets. In particular, this has proven to be invaluable
for $\gamma$\,Dor stars, as their gravity and/or gravito-inertial mode
frequencies form a very dense spectrum in the range of 0.3 to 3\,$\rm
d^{-1}$. Period spacing patterns have now been detected for dozens of
$\gamma$\,Dor stars \citep[e.g.][]{Chapellier2012, Kurtz2014, Bedding2015,
  Saio2015, Keen2015, VanReeth2015,Murphy2016}.

In this study we focus on the period spacing patterns detected by
\citet{VanReeth2015} in a sample of 68 $\gamma$\,Dor stars with spectroscopic
characterisation and aim to derive the stars' internal rotation rate and the
asymptotic period spacing value of the series. This serves as a first step for
future detailed analyses of differential rotation, similar to the studies which
have previously been carried out in slow rotators among g-mode pulsators
  interpreted recently in terms of angular momentum transport by internal
  gravity waves \citep[e.g.,][]{Triana2015,Rogers2015}. In this paper we present
  a grid of theoretical models, which we use as a starting point (Section
  \ref{sec:grid}), and explain our methodology to derive the rotation
    frequency (Section \ref{sec:method}). The
  method is illustrated with applications on synthetic data (Section
  \ref{subsec:synthdata}), a slowly rotating star with rotational splitting,
  KIC\,9751996, and a fast rotator with a prograde and a retrograde period
  spacing series, KIC\,12066947 (Section \ref{subsec:slowfast}). We then analyse
  the sample as a whole (Section \ref{subsec:sample}), before moving on to the
  discussion and plans for future in-depth modelling of individual targets
  (Section \ref{sec:conclusions}).

\section{Grid of stellar models and pulsation frequencies}
\label{sec:grid}
We first computed a rough grid of theoretical stellar models to gain further
insight into the internal structure and properties of $\gamma$\,Dor stars. To
allow for a complete understanding, the models were purposely kept relatively
simple. We did not include any rotational effects into the equilibrium models,
allowing us to assume spherical symmetry and compute 1-dimensional models with
the 1D MESA stellar evolution code
\citep[v7385;][]{Paxton2011,Paxton2013,Paxton2015}. The convection was treated
using the mixing length theory with $\alpha_{\rm MLT} = 1.8$ and the Ledoux
criterion with $\alpha_{\rm sc} = 0.01$. A single diffusive mixing coefficient
was defined in the radiative region and fixed at a value of $1\,cm^2s^{-1}$. We
used the solar metallicity values given by \citet{Asplund2009} and OPAL type I
opacity tables \citep{Rogers2002}. The varying parameter values of the models in
the grid are given in Table \ref{tab:mesa}.

\begin{figure}
 \includegraphics[width=88mm]{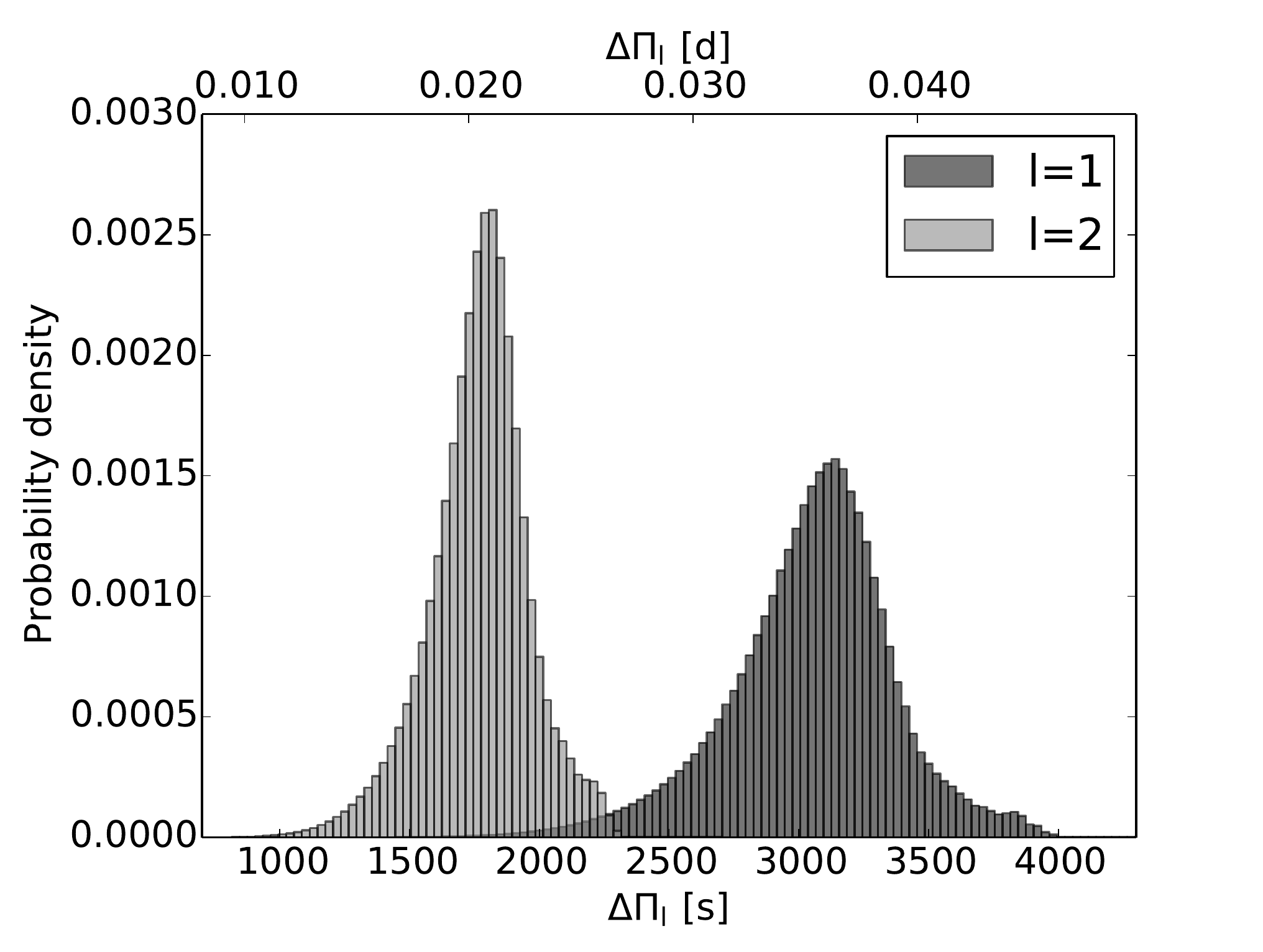}
 \caption{\label{fig:Spacing_dist}The distributions of the asymptotic period
   spacing values $\Delta\Pi_l$ for spherical degree $l=1$ and $2$, computed for
   the MESA evolution tracks with the input parameters provided in Table
   \ref{tab:mesa}. For the computation of the distributions the ages and
   evolution rates of the stellar models were taken into account.}
\end{figure}

For each of the models in our grid we also computed the asymptotic period spacing 
\begin{equation}
\Delta\Pi_l = \frac{\Pi_0}{\sqrt{l(l+1)}},
\label{eq:1}
\end{equation}
with 
\begin{equation}
\Pi_0
= 2\pi^2\left(\int_{r_1}^{r_2}N\frac{\mathrm{d}r}{r}\right)^{-1},
\label{eq:2}
\end{equation}
as derived by \citet{Tassoul1980} for high-order gravity modes. Here $l$ is the
spherical degree of the pulsation mode, $r$ is the distance from the stellar
center, $N$ is the Brunt-V\"ais\"al\"a frequency and the boundaries of the mode
trapping region are marked by $r_1$ and $r_2$ \citep{Aerts2010}. While $\Delta\Pi_l$ is smaller
for larger values of $l$ (Eq.\,\ref{eq:1}), $\Delta\Pi_l$ also changes as the
star evolves. We have therefore calculated the probability of observing
different spacing values using the stellar ages in our grid models. As
shown in Fig.\,\ref{fig:Spacing_dist}, we typically expect $\Delta\Pi_l$ values
on the order of 3100\,$s$ and 1800\,$s$ for $l=1$ and $l=2$ respectively, 
  which in turn implies $\Pi_0$ is 
on the order of $4400\,s$. In
addition, there are strong linear correlations for $\Delta\Pi_l$ between models
with different values of $M$, $Z$, $X$, $f_{\rm ov}$ and $\alpha_{\rm ov}$,
assuming a fixed hydrogen abundance $X_c$ in the convective core.

\begin{figure}
 \includegraphics[width=88mm]{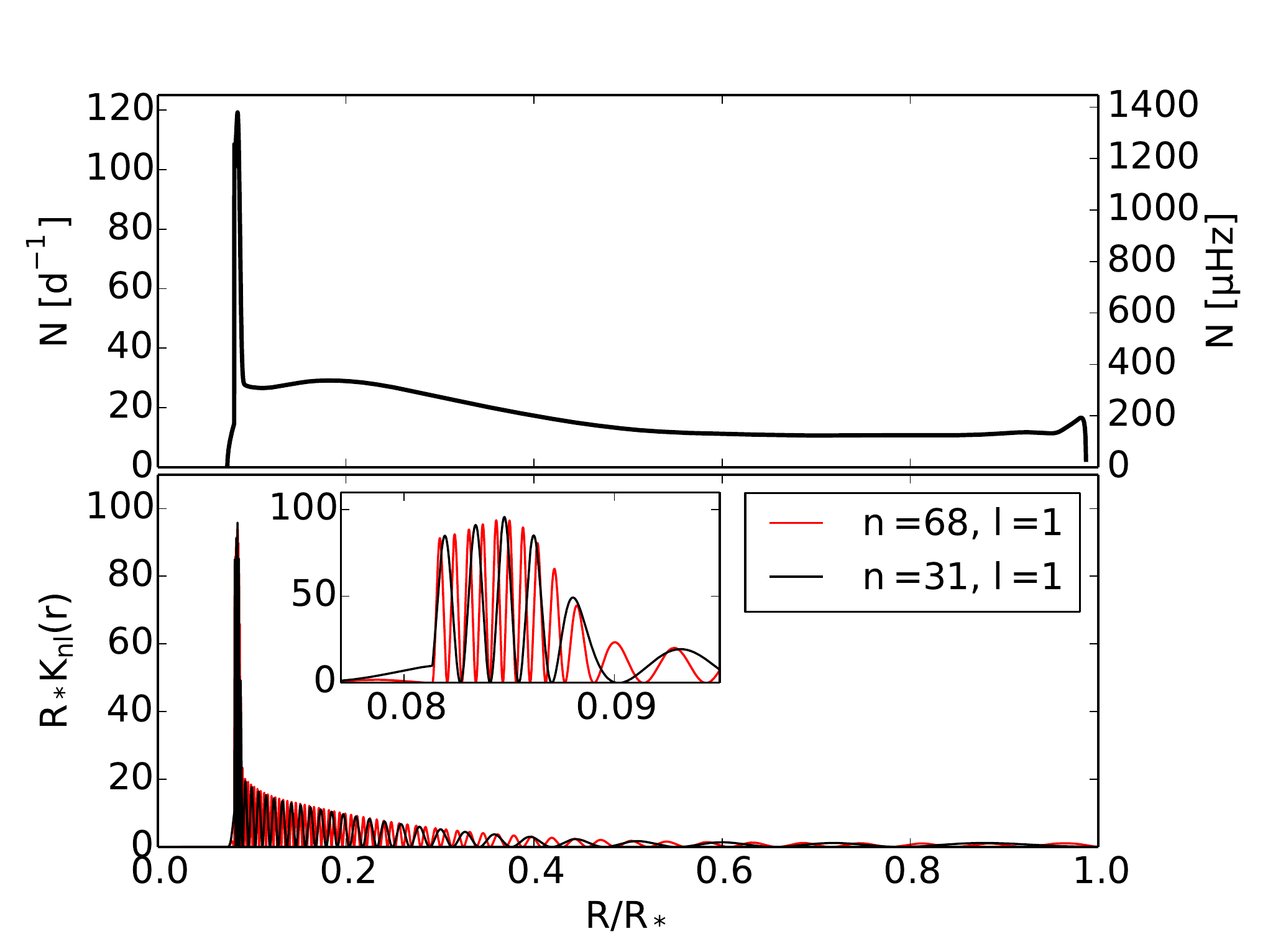}
 \caption{\label{fig:kernels} The Brunt-V\"ais\"al\"a frequency $N$ (\emph{top}) and the rotational kernel $K_{nl}$ (\emph{bottom}) for the lowest- and highest-order mode of the stellar model discussed in Sec.\,\ref{subsec:synthdata}. The inset shows a zoom of $K_{nl}$. Both functions correlate with the sensitivity of the gravity-mode pulsations to the different regions inside the star.}
\end{figure}

As shown by \citet{Bouabid2013} and as observed by \citet{VanReeth2015},
gravity-mode period spacing patterns are heavily influenced by rotation. We
therefore introduced the influence of rotation on the pulsation periods using
the traditional approximation \citep{Eckart1960,Lee1987,Townsend2005}. In this
framework the $\theta$-component of the rotation vector is ignored
\citep{Lee1997} and it is assumed that the star is sufficiently slowly rotating,
so the effects of the centrifugal force can be neglected. While this
  particular assumption may not always be applicable, gravity-modes and
  gravito-inertial modes are mostly sensitive to the stellar properties near the
  convective core, where the rotational deformation of the star remains
  limited. This is illustrated in Fig.\,\ref{fig:kernels}, which shows both the
Brunt-V\"ais\"al\"a frequency $N$ and the rotational kernel $K_{nl}$ for the
lowest- and highest-order mode of the stellar model discussed in
Sec.\,\ref{subsec:synthdata}. Both functions correlate with the sensitivity of
the pulsations to the different regions in the star and peak near the convective
core boundary. The rotational kernel $K_{nl}$ specifically indicates the
sensitivity of the pulsations to the local stellar rotation profile. Thus,
the rotational frequencies deduced from pulsational
  properties throughout this paper correspond with the near-core interior rotation rates.

\citet{Ballot2011} showed that the traditional approximation
  continues to perform adequately if the spin parameter $|s| \leq 2$
  \citep[see][Fig.\,2]{Ballot2011}, with
\begin{equation}s = \frac{2f_{\rm rot}}{f_{co}}, \label{eq:3} 
\end{equation}
where $f_{\rm rot}$ ($=\Omega/2\pi$) and $f_{co}$ are the stellar rotation
frequency and the pulsation frequency in the corotating frame respectively.
Thanks to the assumptions made in the traditional approximation, the
computational requirements for the effects of rotation are dramatically
reduced. In this work, we used the traditional approximation module from the 1D
pulsation code GYRE v4.3 \citep{Townsend2013}, and follow the approach described
by \citet{Ballot2011} and \citet{Bouabid2013}. These authors show that,
  within the traditional approximation, an asymptotic pulsation period series
  can be rewritten for a rotating star as 
\begin{equation}
P_{nlm,co} = \frac{\Pi_0}{\sqrt{\lambda_{l,m,s}}}(n + \alpha_{g}),
\label{eq:4} 
\end{equation}
where $P_{nlm,co}$ is the pulsation period of radial order $n$, spherical degree
$l$ and azimuthal order $m$ in the corotating frame. In this paper, we adopt the
convention that $m >
0$ corresponds with prograde modes and $m < 0$ with retrograde modes,
respectively. In this equation, $\lambda_{l,m,s}$ is the eigenvalue of the Laplace tidal equation
depending on $l$, $m$, and the spin parameter $s$, while the phase term $\alpha_{g}$
depends on the internal stellar properties at the boundaries of the pulsation
mode cavity and can be taken to be 0.5 for stars with a convective core and a
convective envelope, such as $\gamma$\,Dor stars. In the limit of a non-rotating
star, where $s = 0$, this expression reduces to  \begin{equation} P_{nl} =
  \frac{\Pi_0}{\sqrt{l(l+1)}}(n + \alpha_{g}),\label{eq:5} 
\end{equation} in agreement with Eq.\,(\ref{eq:1}) and as derived by
\citet{Tassoul1980}.
For each of the models in our grid, we computed the $l=1$ and $l=2$ mode
frequencies for radial orders ranging from 5 to 120.

 \begin{table} 
\centering
\begin{tabular}{lccc}
\hline\hline
 parameter & begin & end & step size\\
 \hline
 mass $M$ [$M_\odot$] & 1.4 & 2.0 & 0.05\\
 metallicity $Z$ & 0.010 & 0.018 & 0.004\\
 exp. core overshooting $f_{ov}$ & 0.001 & 0.03 & 0.0075\\
 step core overshooting $\alpha_{ov}$ & 0.01 & 0.3 & 0.075\\
 initial hydrogen abundance $X_i$ & 0.69 & 0.73 & 0.02\\
 \hline
  \end{tabular}
\caption{\label{tab:mesa} The parameter values of the computed grid of 1170 MESA
  evolutionary tracks, consisting of some 900,000 models.}
\end{table}

\section{Methodology}
\label{sec:method}
  As we have discussed in the previous section, the influence of rotation
  on the pulsation frequencies depends on the values of both $l$ and $m$, while
  the asymptotic spacing $\Delta\Pi_l$ is dependent of the value of $l$
  (Eq.\,\ref{eq:1}). It is therefore necessary to have a pulsation mode
  identification if we wish to constrain the rotation profile of the observed
  star properly.

\begin{figure}
 \includegraphics[width=88mm]{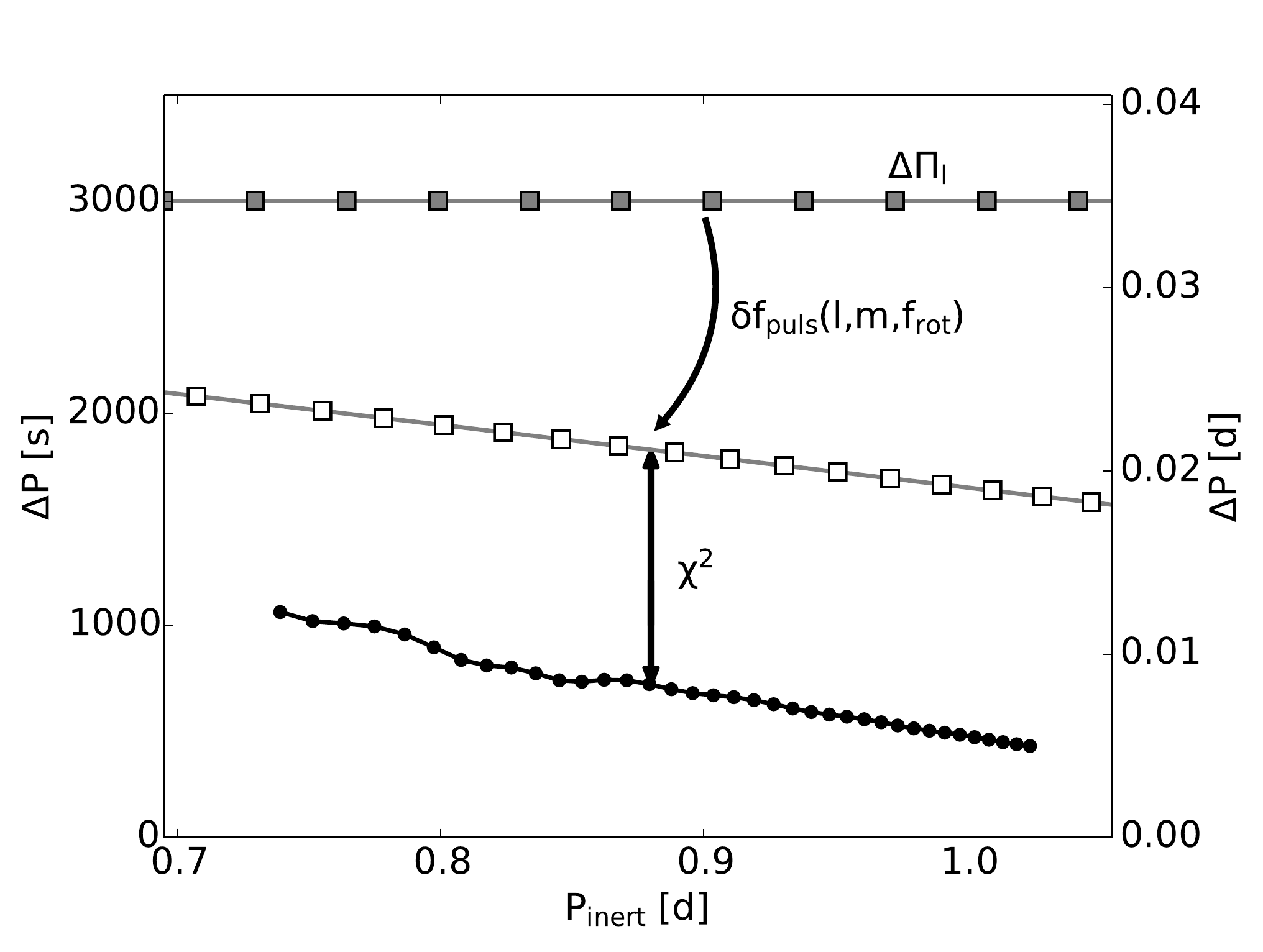}
 \caption{\label{fig:method} Illustration of our methodology to derive the rotation rate $f_{\rm rot}$ and asymptotic spacing $\Delta\Pi_l$ from an observed period spacing pattern (black dots). An equidistant spacing series (grey squares) is defined, rotationally shifted (white squares) and fitted to the observed pattern using $\chi^2$-minimisation, optimising for the variables $l$, $m$, $\Delta\Pi_l$ and $f_{rot}$.}
\end{figure}

To derive a reliable estimate of the rotation rate of a $\gamma$\,Dor
  star with one or more observed period spacing patterns, we consider all the
  possible combinations of $l$ and $m$-values for the mode identification of the
  GYRE pulsation frequencies computed for the MESA models in our grid.  For each
  combination of (l,m), we compute the asymptotic spacing value $\Delta\Pi_l$,
  as expressed in Eq.\,(\ref{eq:1}) and subsequently correct it in the framework
  of the traditional approximation according to Eq.\,(\ref{eq:5}). This is
  illustrated graphically in Fig.\,\ref{fig:method}.  Because the application
of a rotational frequency shift does not introduce dips into the period spacing
patterns, we do not need to take them into account at this point. A uniform
period spacing series is sufficient for our needs.

The pulsation frequencies in this series are then rotationally shifted using the
traditional approximation, as described by Eq.\,\ref{eq:4} and assuming
  the star is rigidly rotating. The values of the pulsation periods in the
inertial reference frame are then obtained by \begin{equation}P_{\rm inert} = \frac{1}{f_{co}
  + mf_{\rm rot}}.\label{eq:6} 
\end{equation} This introduces a slope into the model spacing series, as
shown in Fig.\,\ref{fig:method}. The resulting pattern is subsequently fitted to
the observed period spacing series using $\chi^2$-minimisation,
optimising for the variables $\Delta\Pi_l$ and $f_{rot}$. Finally, we
  select the best solution for all studied $l$ and $m$ values, taking into
  account the theoretical expectations for the asymptotic spacing $\Delta\Pi_l$,
  as shown in Fig.\,\ref{fig:Spacing_dist} and derived from our model grid in
  Sec.\,\ref{sec:grid}. From this fit, we then obtain estimates for the
rotation rate $f_{rot}$ and the asymptotic spacing $\Delta\Pi_l$, as well as a
mode identification.

\section{Applications}
\label{sec:appl}
\subsection{Synthetic data}
\label{subsec:synthdata}
To illustrate our method we first analyse a simulated period spacing
pattern. The simulated data were computed using the MESA and GYRE codes with the
input values provided in Table \ref{tab:synth}, further taking ($l$,$m$)
  = (1,1). For the computation of the evolution track itself the influence of
rotation was not taken into account. The rotation was only included in the GYRE
computations using the traditional approximation module. The computed pattern is
shown in Fig.\,\ref{fig:synth_obsfit} and the values of the pulsation periods
are listed in the appendix in Table \ref{tab:synthP}.

\begin{figure}
 \includegraphics[width=88mm]{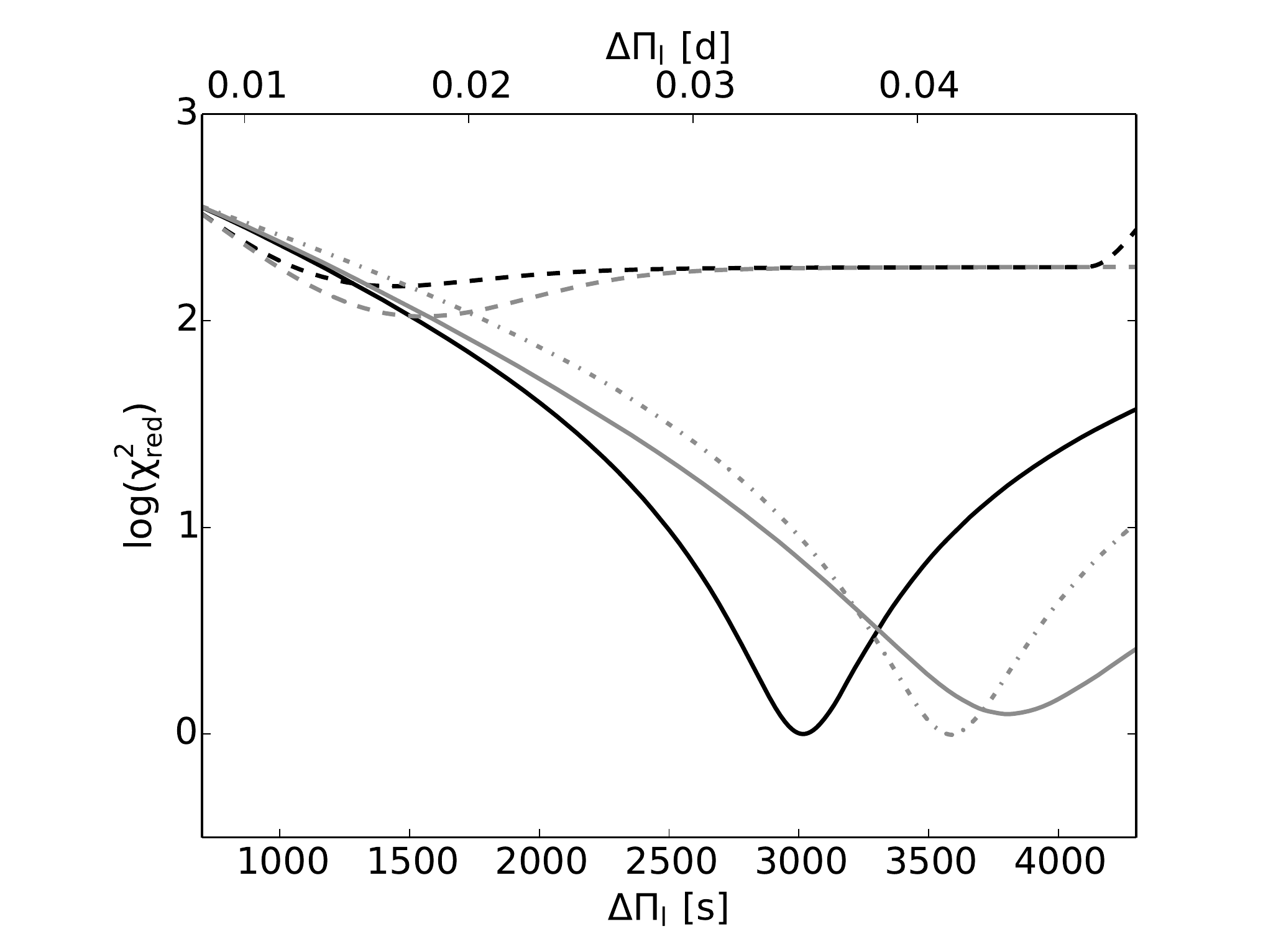}
 \caption{\label{fig:synth_chi2} The best $\chi^2$-values 
for the synthetic data shown in Fig.\,\ref{tab:synth} for each
   ($l,m$)-combination, as a function of the asymptotic spacing
   $\Delta\Pi_l$. The black and light grey lines correspond to $l=1$ and $l=2$
   respectively, while the modes with $m=0$ are indicated with dashed lines,
   $m=1$ with full lines and $m=2$ with the dash-dotted line.}
\end{figure}

\begin{figure}
 \includegraphics[width=88mm]{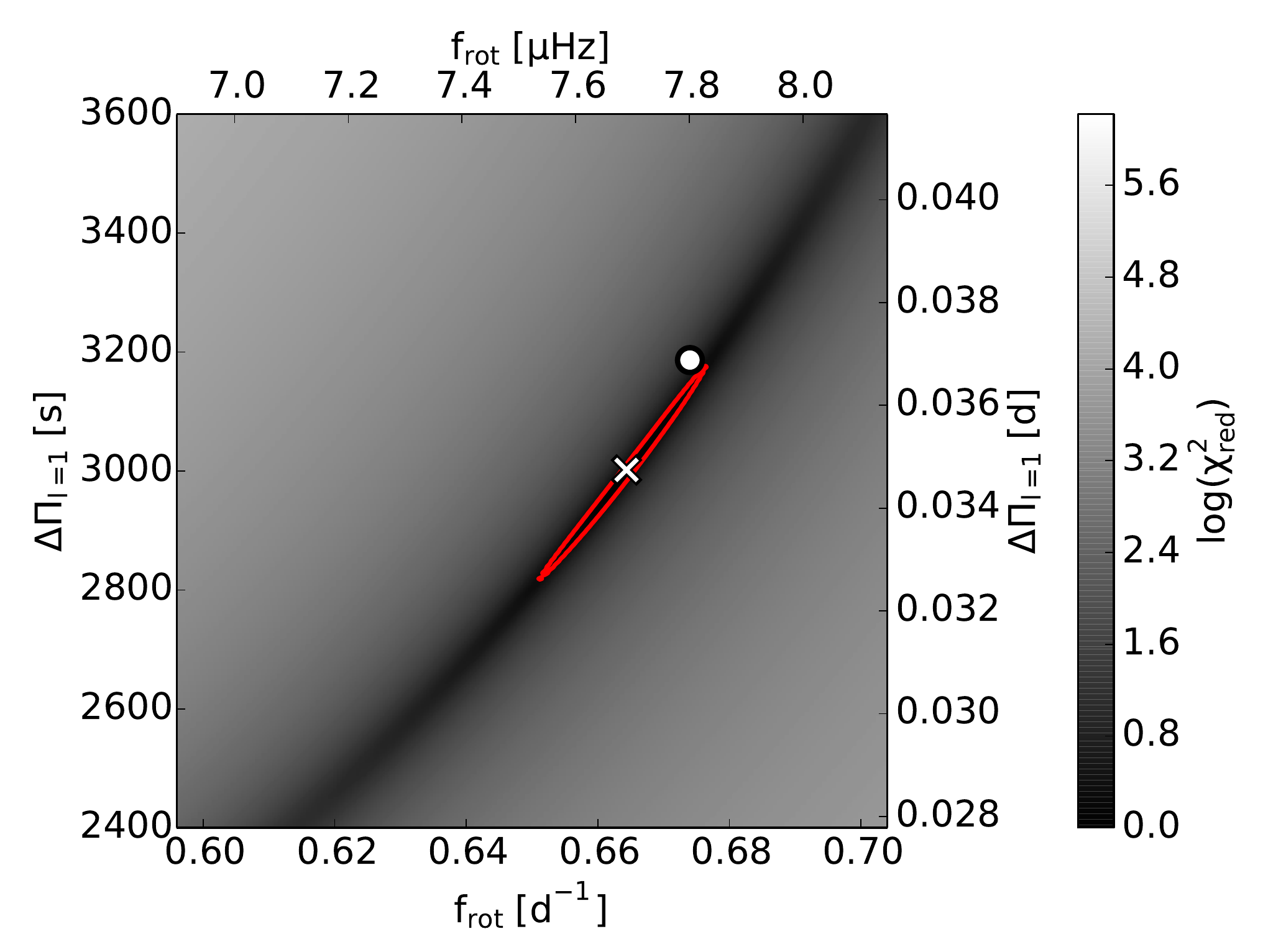}
 \caption{\label{fig:synth_chi2_2d}$\log(\chi^2_{\rm red})$ for the simulated
   period spacing series assuming $l=1$, as a function of the asymptotic period
   spacing $\Delta\Pi_{l}$ and the rotation rate $f_{rot}$. The white dot
   indicates the input values, while the white cross and the red boundary
   indicate the found solution and its 1$\sigma$-uncertainty margins
   respectively.}
\end{figure}

The results of our analysis are shown in Figs.\,\ref{fig:synth_obsfit} to
\ref{fig:synth_chi2_2d}. As we can see in Fig.\,\ref{fig:synth_obsfit} we fitted
the simulated data nicely when we excluded the dips in the pattern from the
analysis and assumed ($l$,$m$) = (1,1). However, similar good results
were obtained when we treated the pulsations as (2,1)-modes or (2,2)-modes
during our analysis, as illustrated in Fig.\,\ref{fig:synth_chi2}. In other
words, we cannot obtain a clear mode identification if we base ourselves solely
on the obtained $\chi^2$-values. This problem is solved when we look back to the
expected values of the asymptotic spacing $\Delta\Pi_l$ for different values of
$l$, which we previously showed in Fig.\,\ref{fig:Spacing_dist}. It is clear
that the found values for $\Delta\Pi_l$ are far too large for $l=2$. We can
therefore safely identify the simulated data as (1,1)-modes, and obtain $f_{\rm
  rot} = 0.664\pm0.013\,d^{-1}$ and $\Delta\Pi_{l=1} = 3020\pm190\,s$.

It is important to exclude any significant dips in the period spacing structure
from this analysis. In our technique, we do not take the influence of chemical
gradients in the stellar interior into account. As shown by \citet{Miglio2008a},
these result in non-uniform deviations from the asymptotic spacing
series. Because we can only observe a small part of a period spacing pattern,
any non-uniform variations in the pattern will change the measured mean spacing
and/or the measured slope of the pattern. This, in turn, will influence our
analysis. By ignoring significant non-uniform variations in the period spacing
structure, we limit their influence on the analysis, so that we obtain results
which are correct within or on the order of $1\sigma$. For our simulated data
set this is illustrated with the 2-dimensional $\chi^2$-distribution shown in
Fig.\,\ref{fig:synth_chi2_2d}. While we ignored the large dip in the period
spacing structure (as seen in Fig.\,\ref{fig:synth_obsfit}), the remaining
non-uniform variations still impacted the analysis. As a result, there is a
small offset between the input values of the data set and the
1$\sigma$-confidence interval for the obtained solution.

 \begin{table}
\centering
\begin{tabular}{lc}
\hline\hline
parameter & values\\
\hline
 mass $M$ [$M_\odot$] & 1.63\\
 metallicity $Z$ & 0.016\\
 initial hydrogen abundance $X_i$ & 0.71\\
  mixing length parameter $\alpha_{MLT}$ & 1.8  \\
  step core overshooting $\alpha_{ov}$ & 0.18\\
  mixing coefficient $D$ [$cm^2 s^{-1}$] & 0.8 \\
  \hline
  $T_{\rm eff}$ [K] & 7047\\
  $\log g$ [dex] & 4.34\\
  $[M/H]$ [dex] & 0.094\\
  $v_{\rm eq}$ [km\,$\rm s^{-1}$] & 69.38\\
  $f_{\rm rot}$ [$\rm d^{-1}$] & 0.674\\
  $\Delta\Pi_{l=1}$ [$s$] & 3186.5\\
  central hydrogen abundance $X_c$ & 0.357\\
  \hline
\end{tabular} 
\caption{\label{tab:synth} The parameter values of the simulated period spacing
  pattern. \emph{Top:} the input parameters of the MESA evolution
  track. \emph{Bottom:} the parameters of the model for which the pulsation
  periods were computed.}
\end{table}

\begin{figure}
 \includegraphics[width=88mm]{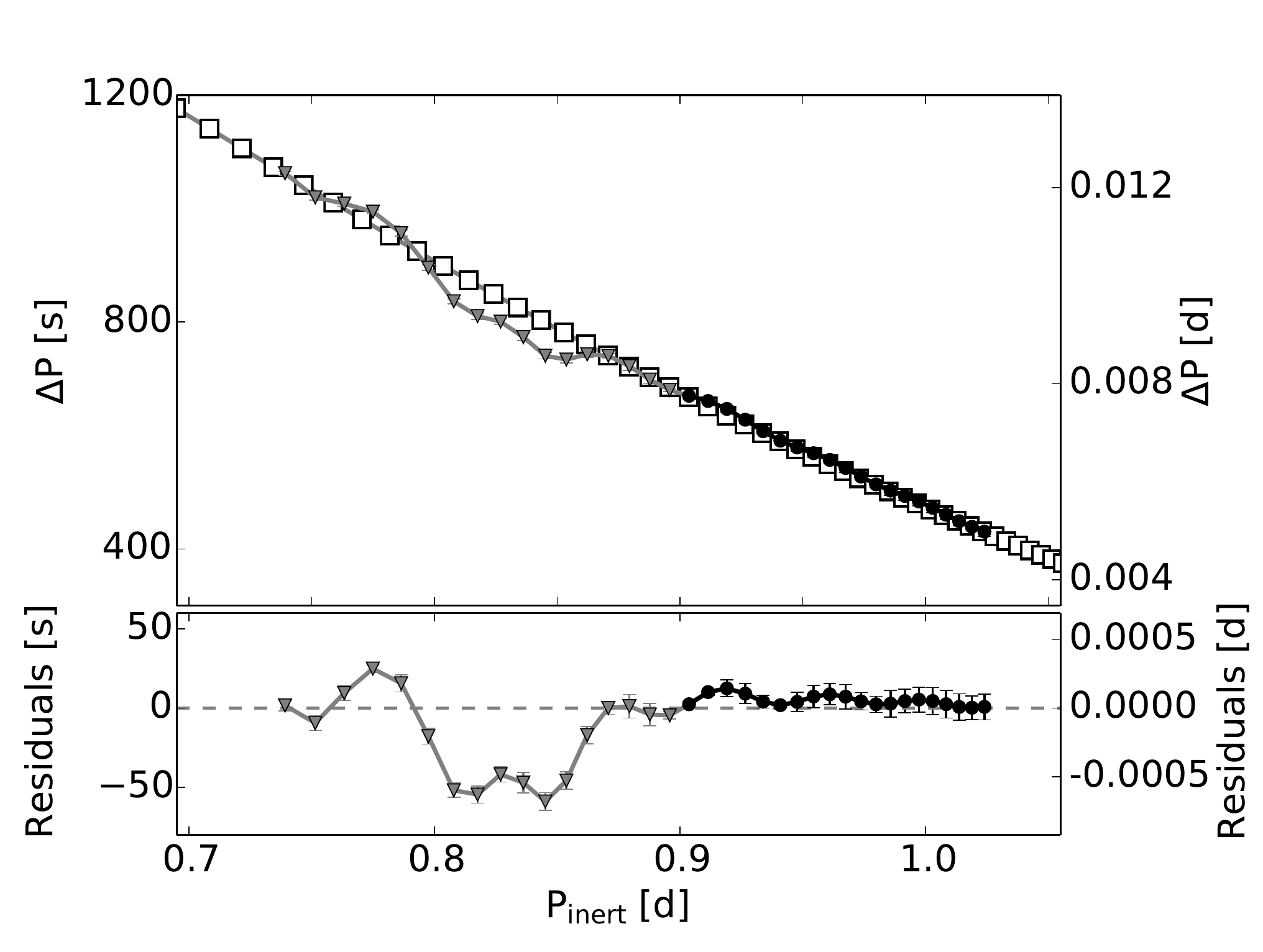}
 \caption{\label{fig:synth_obsfit} \emph{Top:} the input period spacing pattern
   (black dots and grey triangles) with the best-fitting pattern (white squares)
   as obtained from the $\chi^2$-minimisation in Fig.\ref{fig:synth_chi2_2d}
   assuming ($l$,$m$) = (1,1). The black part of the input patterns was
   used to determine $f_{rot}$ and $\Delta\Pi_{l}$, while the grey section was
   excluded. \emph{Bottom:} the residuals of the fit.}
\end{figure}

\subsection{A slow and a fast rotator}
\label{subsec:slowfast}
\begin{figure}
 \includegraphics[width=88mm]{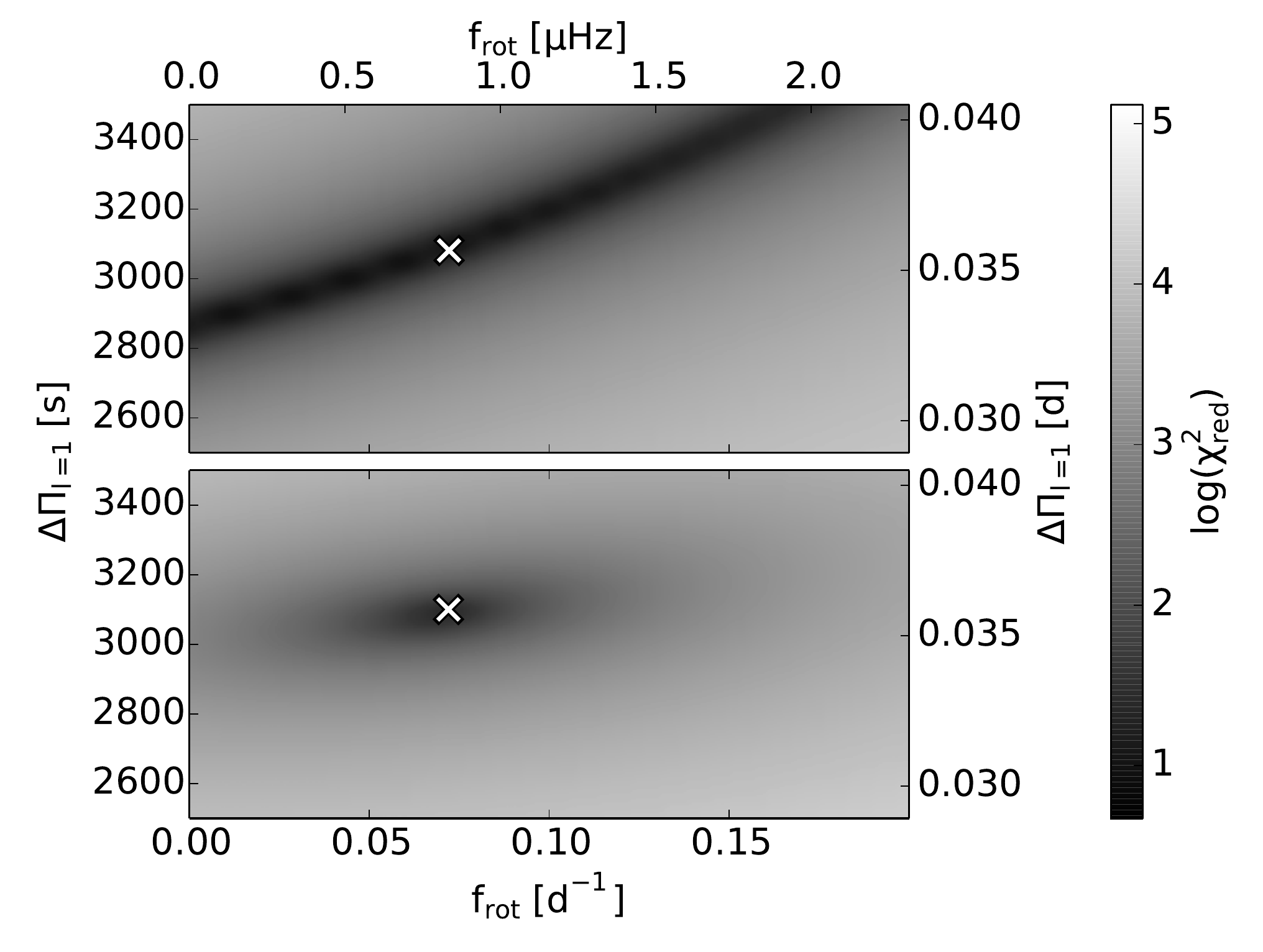}
 \caption{\label{fig:kic9751996_chi2comp} $\log(\chi^2_{\rm red})$ for the
   observed period spacing series of KIC\,9751996, assuming $l=1$, as a function
   of the asymptotic period spacing $\Delta\Pi_{l}$ and the rotation rate
   $f_{rot}$. The white crosses indicate the optimal found
   solutions. \emph{Top:} the $\chi^2$-distribution that we find by only
   analysing the detected prograde series. \emph{Bottom:} the
   $\chi^2$-distribution obtained by fitting the prograde, zonal and retrograde
   spacing series simultaneously.}
\end{figure}

\begin{figure}
 \includegraphics[width=88mm]{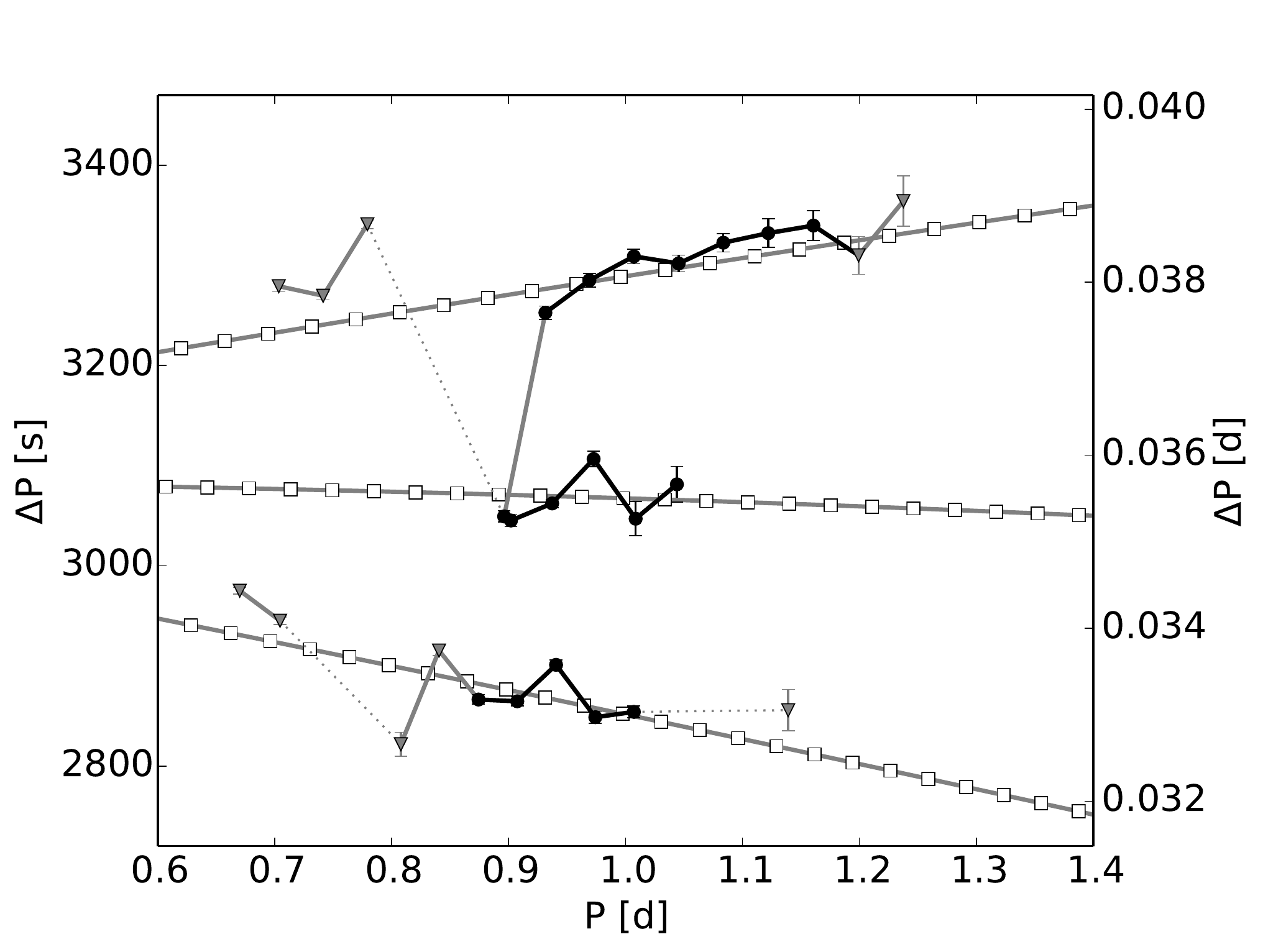}
 \caption{\label{fig:kic9751996_fit} The observed period spacing patterns (black
   dots and grey triangles) for the retrograde (top), zonal (middle) and
   prograde (bottom) modes of KIC\,9751996. The black parts of the input
   patterns were used to determine $f_{rot}$ and $\Delta\Pi_{l}$, while the grey
   sections were excluded. The white squares indicate the modes of the
     optimal model in the grid when all three series are fitted simultaneously.}
\end{figure}

In our sample we have one slowly rotating star, KIC\,9751996, for which we
detected period spacing series with rotational splitting, delivering
  immediately the $m$-values of the  modes. In order to further
validate our proposed methodology, we have applied it to KIC\,9751996. In a
first step, we only analysed the prograde period spacing pattern to test the
reliability of our method. Assuming ($l$,$m$) = (1,1), this led us to
find $f_{\rm rot} = 0.07\pm0.02\,d^{-1}$, which is shown in the top of
Fig.\,\ref{fig:kic9751996_chi2comp}. However, assuming $l=1, m=0$ for the
  treated series, we found
$f_{\rm rot} = 0.19\pm0.03\,d^{-1}$ for a similar $\chi^2$ value. The challenge
in this case is that the shift and the slope in the period spacing pattern are
almost negligibly small compared to the non-uniform period spacing variations
due to a chemical gradient. This was resolved when we fit the prograde, zonal
and retrograde dipole modes simultaneously, as shown in
Fig.\,\ref{fig:kic9751996_fit}. Not only did this allow us to formally identify
the $(l,m,n)$-values of the
modes, it also resulted in a much higher precision for the rotation rate
$f_{\rm rot} = 0.0696\pm0.0008\,d^{-1}$ and the spacing $\Delta\Pi_{l=1} =
3086\pm6\,s$ (See Fig.\,\ref{fig:kic9751996_chi2comp}). Interestingly,
  we have another indication for 
this derived rotation rate independently. Both in the series
  of the prograde modes and of the retrograde modes, we have a pulsation period
  which does not seem to follow the pattern, at values of 0.8 days and 0.9 days,
  respectively. These modes are likely trapped, which has influenced their
  pulsation period. When the periods of these retrograde and prograde modes are
  converted to their values in the corotating reference frame using the derived
  rotation rate, we find the pulsation periods are almost equal, which is
  consistent with the interpretation of trapped pulsation modes.

\begin{figure}
 \includegraphics[width=88mm]{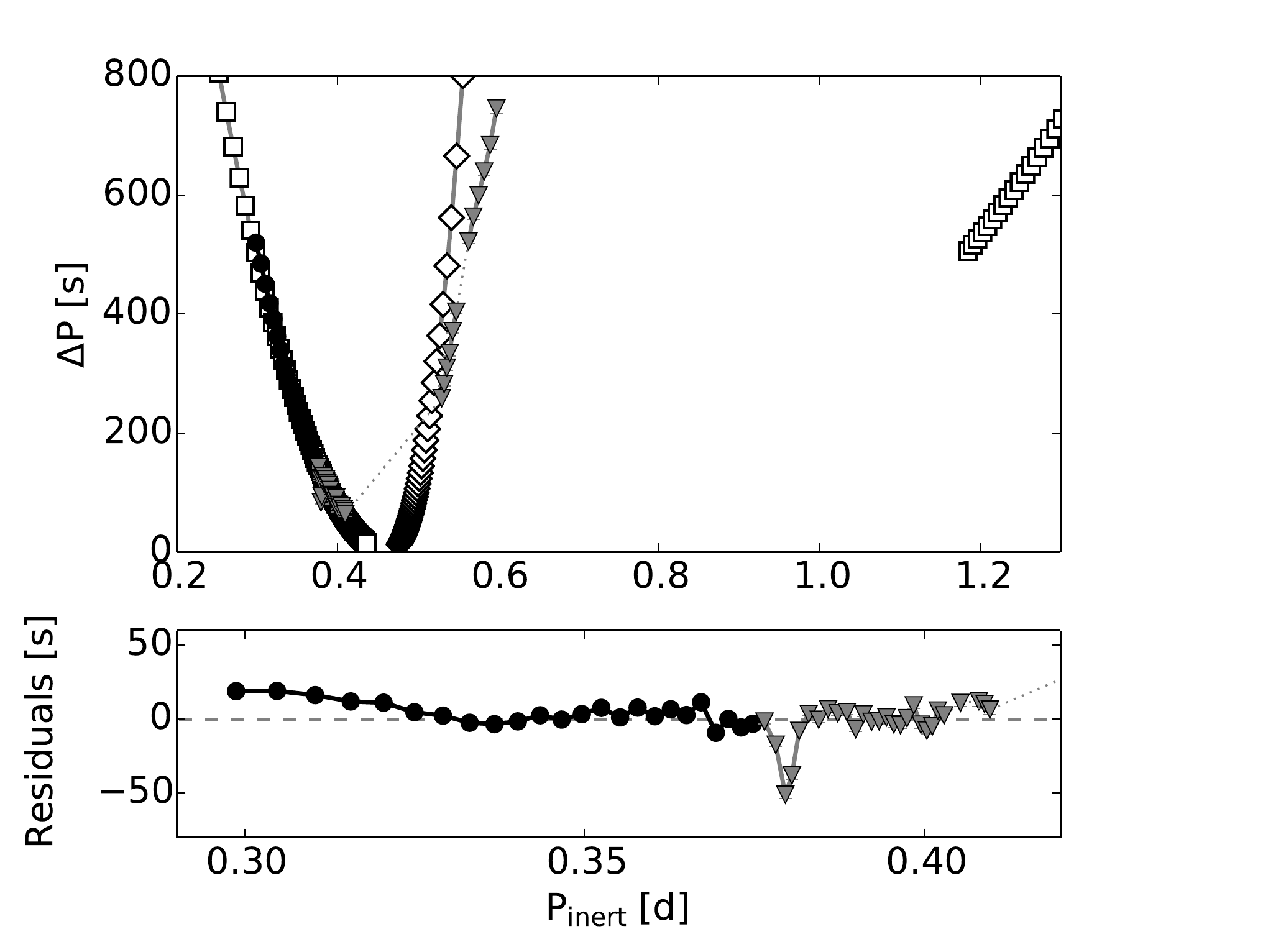}
 \caption{\label{fig:kic12066947_retro} \emph{Top:} The observed period spacing
   patterns (black dots and grey triangles) for the prograde (left) and
   retrograde (right) modes of KIC\,12066947. The black part of the pattern was
   used to determine $f_{rot}$ and $\Delta\Pi_{l}$, while the grey section was
   excluded. The white squares indicate the model fit, assuming we are dealing with gravito-inertial modes and using the $f_{rot}$
   and $\Delta\Pi_{l}$ values obtained from the analysis of the marked prograde
   series. The white diamonds indicate the model computed for Rossby waves using the same $f_{rot}$
   and $\Delta\Pi_{l}$ values, assuming $m=-1$ and $k=-2$ in the k-based
   indexing scheme by \citet{Lee1997}. \emph{Bottom:} The residuals of the fit to the prograde period
   spacing series.}
\end{figure}

Next we also analysed the period spacing patterns of KIC\,12066947, a fast
rotating star for which both a prograde and a retrograde period spacing series
were detected. While we were able to fit the pattern of prograde modes to derive
a rotation rate $f_{rot}$, the observed retrograde series presented us with
  a challenge. We found these to correspond with Rossby modes rather than
  ``classical'' gravity or gravito-inertial modes. Rossby modes can only occur
  in rotating stars and originate from the interaction
  between the stellar rotation and toroidal modes
  \citep[e.g.][]{Papaloizou1978,Townsend2003c}. Our identification of the
  retrograde modes as Rossby modes is illustrated in the top panel of
  Fig.\,\ref{fig:kic12066947_retro}, where we show the observed period spacing
  patterns for both the detected prograde and retrograde series, as well as the
  spacings predicted by the most suitable model in the grid, by assuming the
  values for $f_{rot}$ and $\Delta\Pi_{l}$ obtained by modelling the prograde
  series. For the calculation of the period series in the case of Rossby modes, $\Pi_0$ was derived from
  $\Delta\Pi_{l}$ using Eq.\,(\ref{eq:1}), and this value was subsequently filled into
  Eq.\,(\ref{eq:4}). The appropriate eigenvalues $\lambda$ were computed using the
  asymptotic approximation derived by \citet{Townsend2003c}, i.e. Eq.\,(37) in
  that study. This equation is valid when $\lambda \neq m^2$, as is the case here.
The expected values of $\lambda$ for Rossby modes are three to four
  orders of magnitude smaller than for retrograde gravito-inertial modes, which
  allowed us to identify the observed pulsations. However, as
  \citet{Townsend2003c} pointed out, the asymptotic approximation does not converge well to
  the numerical solution in the case of such modes. The possibility to compute Rossby modes
  has currently not yet been included in the publicly available version of 
GYRE. As a consequence, we could not do
  a reliable quantitative analysis of the retrograde series at this point and
  limited ourselves to the analysis of the prograde series to derive $f_{\rm rot}$. However, several
  qualitative arguments can be made in favour of Rossby modes as a correct identification. From
  the upward slope and the small average period spacing of the observed pattern,
  we derive that these modes are retrograde in the corotating
  frame with $|f_{co}| < f_{rot}$, which is completely in line with the
  theoretical expectations. Furthermore, the observed spin parameter values are
  larger for the retrograde than for the prograde modes, e.g., the values of the
  dominant modes of both series are $15.8\,\pm0.4$ and $7.7\,\pm0.1$,
  respectively. This indicates that in the corotating frame the pulsation
  frequencies of the retrograde modes are smaller than those of the prograde
  modes, which in turn can be explained by the small values of the eigenvalues
  $\lambda$. Finally, \citet{Townsend2003c} also notes that, compared to the
  retrograde gravito-inertial modes, Rossby modes are less equatorially confined as
  the stellar rotation rate increases. As a result, the latter can be expected
  to be less influenced by the geometrical cancellation effects, though the
  effect is still present. For KIC\,12066947, we find that the dominant prograde
  and Rossby modes are confined within equatorial bands with a width of
  77.2\textdegree\ and 53.5\textdegree\ respectively.
  
  For a fast rotating star such as KIC\,12066947, we also have to take into acccount rotational deformation. The centrifugal force leads to a lower effective gravity at the equator than at the pole. This influences the
  Brunt-V\"ais\"al\"a frequency, which affects the pulsations. In the case of KIC\,12066947, we could roughly estimate the
deformation of the star, using equation A.6 from \citet{Maeder2000}. In this
analysis we evaluated the observed spectroscopic parameter values and asymptotic
spacing $\Delta\Pi_{l}$ using the models in our MESA grid and took the best
matching model ($M = 1.5M_\odot$, $Z = 0.014$, $X_i=0.69$, $X_c = 0.452$,
$\alpha_{\rm ov} = 0.01$) as a guess for the stellar structure. We found that
$f_{\rm rot}/f_{\rm rot,crit} = 0.78$ and $R_{\rm pole}/R_{\rm eq} = 0.88$. A
two-dimensional treatment of the rotation is clearly needed to quantify the
impact of the rotation on the  modes, which will allow us to improve
our constraints on $f_{\rm rot}$ and $\Delta\Pi_l$.

\subsection{Sample study}
\label{subsec:sample}
Subsequently, we also applied our methodology to the other stars in our
sample. This led to the mode identification and the determination of the
rotation rate $f_{\rm rot}$ for the period spacing series of 40 stars in our
sample. Six additional sample stars only exhibit retrograde modes and fast
rotation, and cannot be quantitatively analysed with our current
  methodology.  In the case of the remaining four stars, the
difference of the best $\chi^2$-value for different ($l$,$m$) combinations is
too small, so no unique solution could be determined.

For the 40 stars which were successfully analysed, the results are listed in the
appendix in Table \ref{tab:param}. The vast majority of the studied stars were
found to exhibit prograde dipole modes. For fourteen targets in the sample we
had detected multiple series. In principle, these are prime targets to look for
differential rotation. However, for ten of them the second detected period
spacing pattern corresponds to Rossby modes, for which we still need to
  develop a suitable computational tool to arrive at appropriate numerical values,
as discussed in Section \ref{subsec:slowfast}. For the remainder of this study,
we assign the values of $f_{\rm rot}$ and $\Delta\Pi_l$ which we obtained from
the prograde series to the retrograde series of the same star. Because formal
mode identification of these retrograde pulsations is currently not possible,
they are marked as ``R'' in Table \ref{tab:param}. For two other stars we have
both a zonal and prograde dipole series, while for a third we have prograde
dipole and quadrupole modes. Finally, KIC\,9751996, the slowly rotating star we
discussed in Sec.\,\ref{subsec:slowfast}, is the only target for which we have a
series of rotationally split multiplets. For each of these last four stars, we
were able to use the multiple detected period spacing patterns to refine the
obtained $f_{\rm rot}$ and $\Delta\Pi_l$.

\begin{figure}
 \includegraphics[width=88mm]{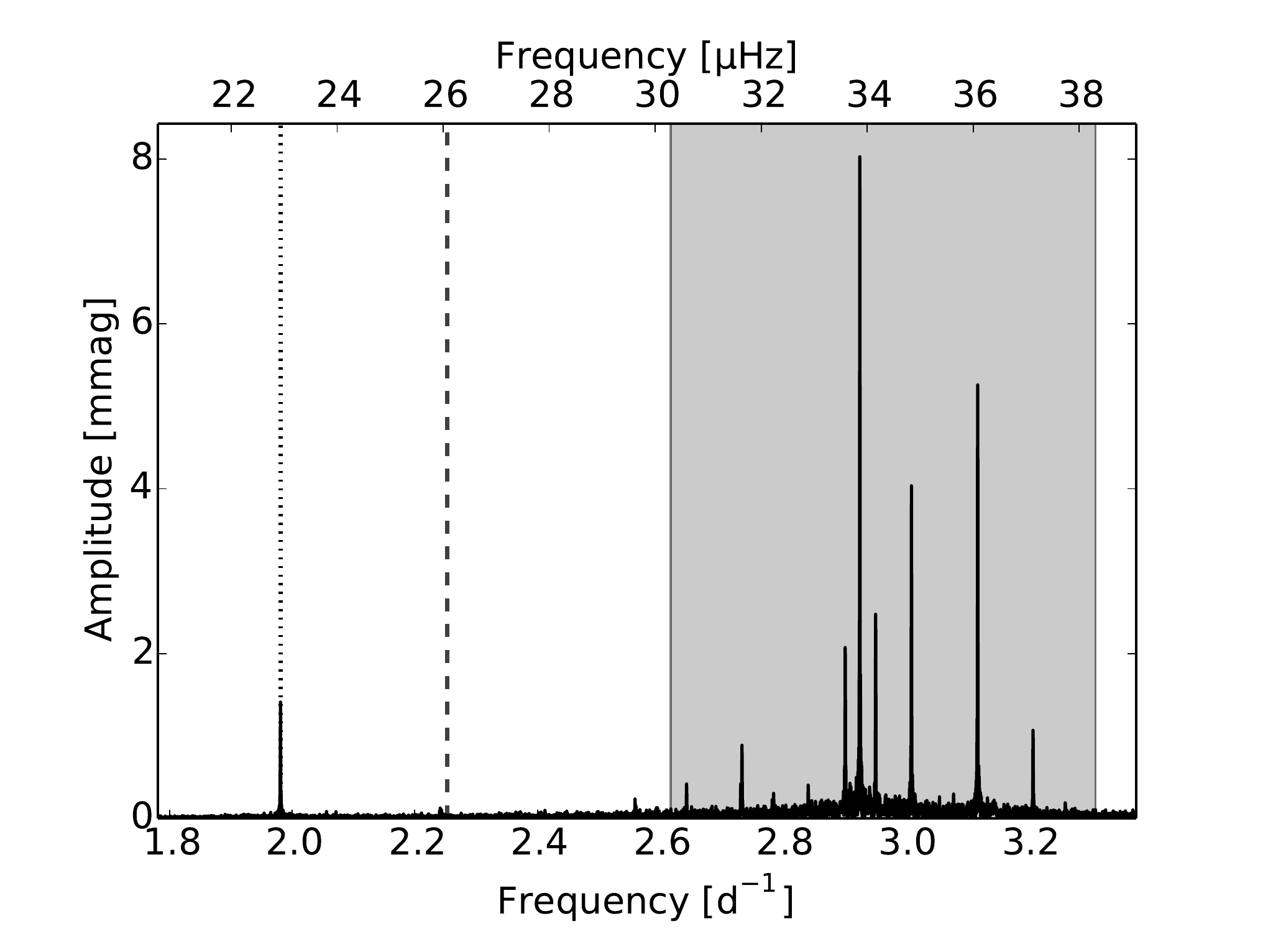}
 \caption{\label{fig:single} Part of the frequency spectrum of KIC\,7365537. The
   light grey area shows the location of the pulsation modes which form the
   detected period spacing pattern of this star. The dashed line marks the value
   of the derived rotation frequency $f_{\rm rot}$, while the dotted line
   indicates the solitary high-amplitude mode which was found.}
\end{figure}

There are several stars for which a single high-amplitude mode was
detected, which does not belong to a period spacing series and which differs
from the rotation frequency. In Fig.\,\ref{fig:single} we show the frequency
spectrum of KIC\,7365537 as an example. For these modes the identification in
Table \ref{tab:param} is marked ``S''. Because our method can not be applied to
these single modes, we again use the values of $f_{\rm rot}$ and $\Delta\Pi_l$
which were derived from the prograde series in the same star, in the
  subsequent analysis. The selection of series of modes in some stars
versus the presence of single modes in others also tells us a great deal about
their respective stellar structure. It has been suggested by
\citet{Dziembowski1991} that such single high-amplitude modes could occur due to
mode trapping effects. However, detailed theoretical modelling of each of these
individual stars is required to confirm this.

\begin{figure*}
 \includegraphics[width=\textwidth]{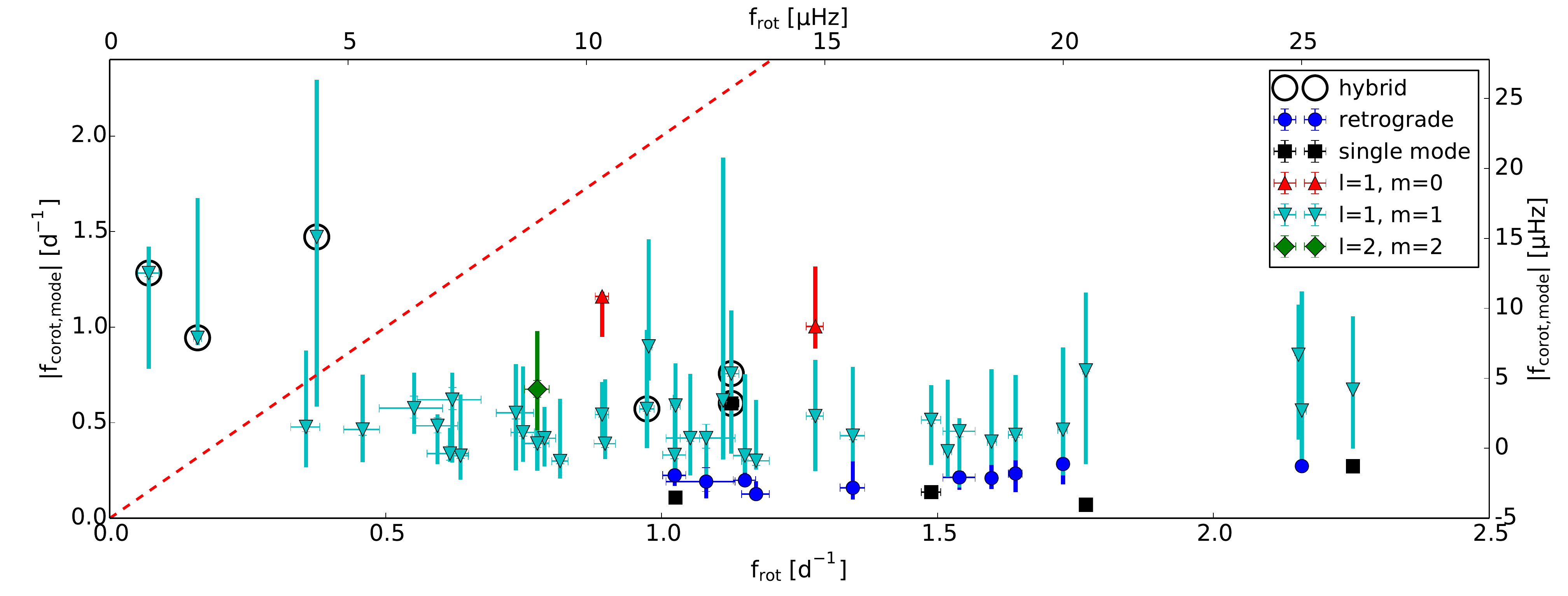}
 \caption{\label{fig:sample_ov} The dominant pulsation frequency $f_{dom,corot}$
   in the corotating frame as a function of the rotation frequency $f_{rot}$ for
   the identified g-mode pulsations of 40 stars in the sample. The thick
   vertical lines indicate the full extent of the detected spacing series. The
   dashed red line marks where the pulsations pass from the superinertial regime
   (above the line) into the subinertial regime (below the line).}
\end{figure*}

\begin{figure*}
 \includegraphics[width=\textwidth]{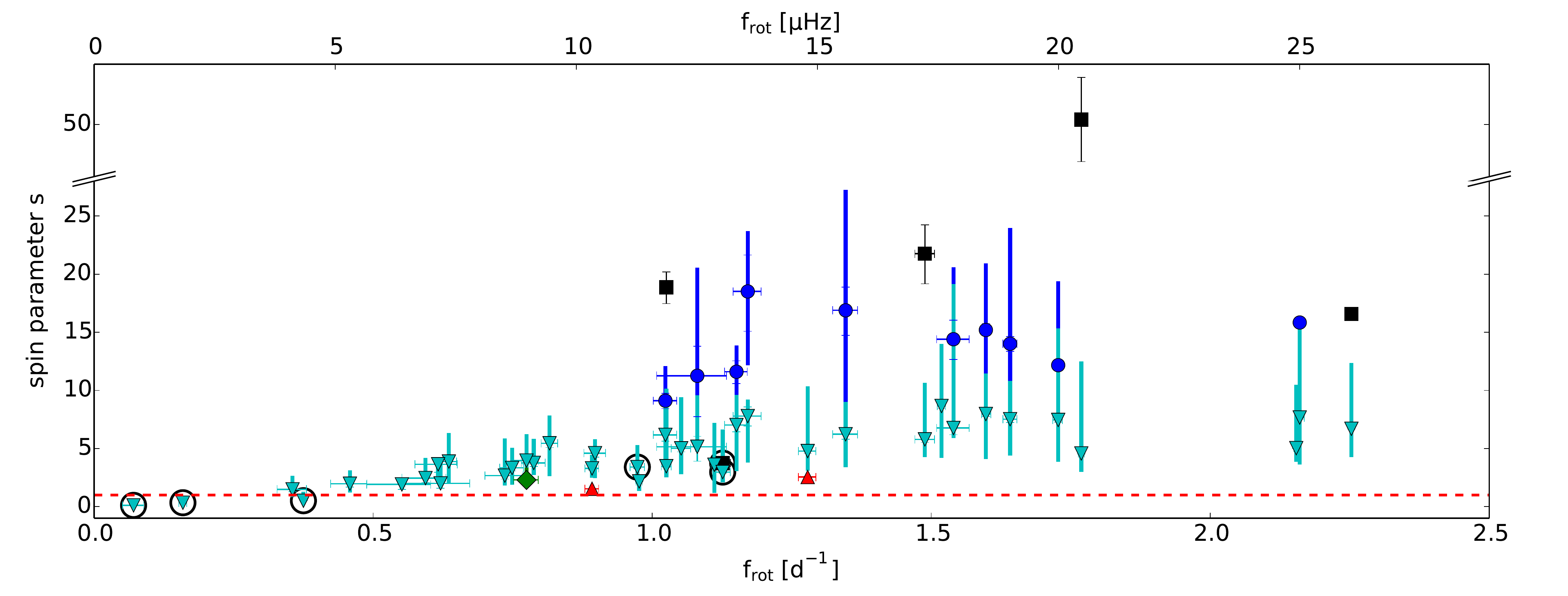}
 \caption{\label{fig:spin} The absolute value of the spin parameter $s$ for the
   detected period spacing series of the stars in our sample, as a function of
   the rotation frequency $f_{rot}$. The dashed red line marks where the
   pulsations pass from the superinertial regime (below the line) into the
   subinertial regime (above the line). The used symbols are the same as in
   Fig.\,\ref{fig:sample_ov}.}
\end{figure*}

Fig.\,\ref{fig:sample_ov} illustrates the frequency $f_{\rm dom,corot}$ of the
dominant mode of each detected series in the corotating frame with respect to
the computed rotation frequency.  An alternative version of
Fig.\,\ref{fig:sample_ov} in the inertial reference frame is included in the
appendix in Fig.\,\ref{fig:sample_inext}.  For the majority of the stars, we
obtain similar values of $f_{\rm dom,corot}$ between 0.15 and 0.75\,$\rm
d^{-1}$. This can be linked to the convective flux blocking excitation
mechanism.  \citet{Dupret2005} and \citet{Bouabid2013} remarked that, in order
for the mode excitation mechanism to be efficient, the thermal timescale
$\tau_{th}$ at the bottom of the convective envelope has to be on the order of
the pulsation periods in the corotating frame. From this information and the
content of Fig.\,\ref{fig:sample_ov}, we can then also derive that both the
detected retrograde spacing series and single modes likely have azimuthal
  order $m = -1$, because only $|m|=1$ led to similar $f_{\rm dom,corot}$
values for the series of different stars. While these results are consistent, we
note that the observed pulsation periods in the corotating frame are typically
larger than the theoretical values computed by \citet{Bouabid2013}. For the
  retrograde Rossby modes this can be linked to the correspondingly low 
  eigenvalues $\lambda$ of the Laplace tidal equation. However, the same
  discrepancy is observed for the prograde modes as well, though to a lesser
  degree. This discrepancy may point towards limitations of the current theory
of mode excitation in $\gamma$\,Dor stars for moderate to fast rotators or
  may be caused by the limited applicability of the traditional approximation
  for these rotation rates. Further research on this topic is required.

In Fig.\,\ref{fig:spin} we show the spin parameter $s$, as defined in
Eq.\,(\ref{eq:3}) and listed in the last column of Table \ref{tab:param}, as a
function of the measured rotation frequency $f_{\rm rot}$. The spin parameter
$s$ is a measure of the impact of rotation on the pulsation frequency and 
  is inversely proportional to the pulsation frequency $f_{\rm co}$ in the
  corotating frame. Once again, the Rossby modes (marked with dark
  blue dots) have small values for the eigenvalue $\lambda$. In addition,
  despite the fact that both prograde sectoral modes and Rossby modes are less easily
  confined in a band around the equator, the effect is still significant for
  these high values of $s$ \citep{Townsend2003c}. This implies many of the stars
  in our sample are seen at moderate to high inclination angles. We further
note that our observed values of $s$ are on average much larger than the values
quoted in theoretical papers in the literature
\citep[e.g.][]{Townsend2003b,Ballot2011}.
  
\begin{figure}
 \includegraphics[width=88mm]{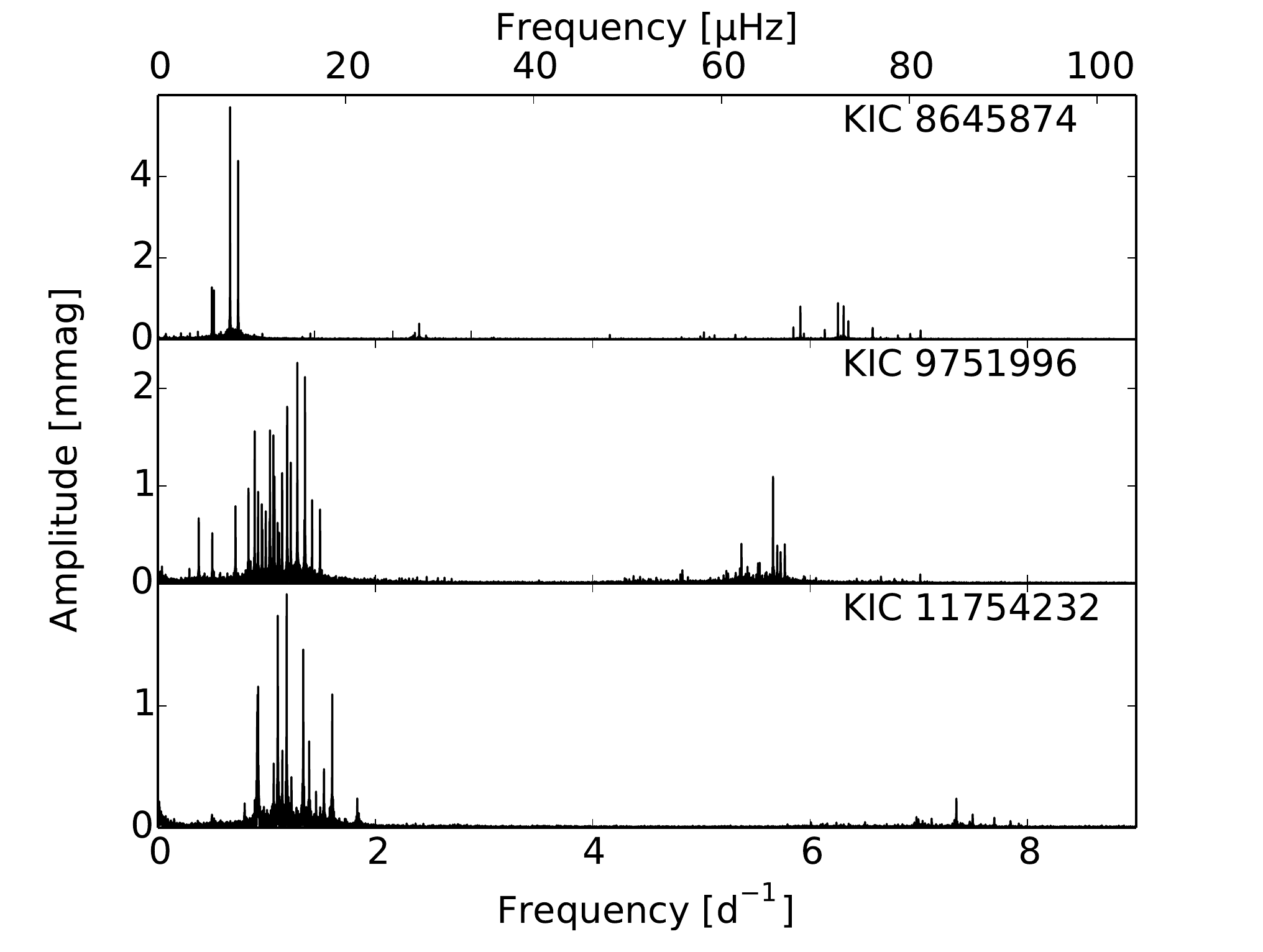}
 \caption{\label{fig:superinert} A section of the Fourier spectra of the three
   slowest rotating stars in our sample. These are our only stars in the
   superinertial regime, and all three are hybrid $\gamma$\,Dor/$\delta$\,Sct
   pulsators with variability between 5\,$\rm d^{-1}$ and 8\,$\rm d^{-1}$.}
\end{figure}

  Fig.\,\ref{fig:superinert} offers a closer look at the three slowest rotating
  stars in our sample, one of which is KIC\,9751996 already discussed in
  Sec.\,\ref{subsec:slowfast}. These three stars have comparable
  properties. They are slow rotators, placing them in the superinertial regime,
  and they are hybrid $\gamma$\,Dor/$\delta$\,Sct pulsators. Each of them
  exhibits variability in the frequency range between 5\,$\rm d^{-1}$ and
  8\,$\rm d^{-1}$. These striking similarities suggest there is a link between
  the stars' low rotation rates and their hybrid properties, marking them as
  interesting targets for follow-up research.

\subsubsection{Statistical analysis}
\label{subsubsec:statan}
Finally, we also look for correlations between the parameter values of our
stars, similar to the multivariate statistical analysis which was carried out by
\citet{VanReeth2015}. In this work, we again use the spectroscopic fundamental
parameter values obtained by \citet{VanReeth2015} in our analysis. We also
include the detected values of the variables $f_{\rm rot}$, $\Delta\Pi_l$,
$R\sin\,i = v\sin\,i/f_{\rm rot}$ and the dominant pulsation frequency $f_{\rm
  dom}$ (both in the corotating and the inertial reference frame). For
consistency, we limit ourselves to the parameter values derived from the
identified prograde dipole mode series of 40 stars in the sample. The results of
our multivariate statistical study are summarised in Table\,\ref{tab:correll}.

\begin{figure}
 \includegraphics[width=88mm]{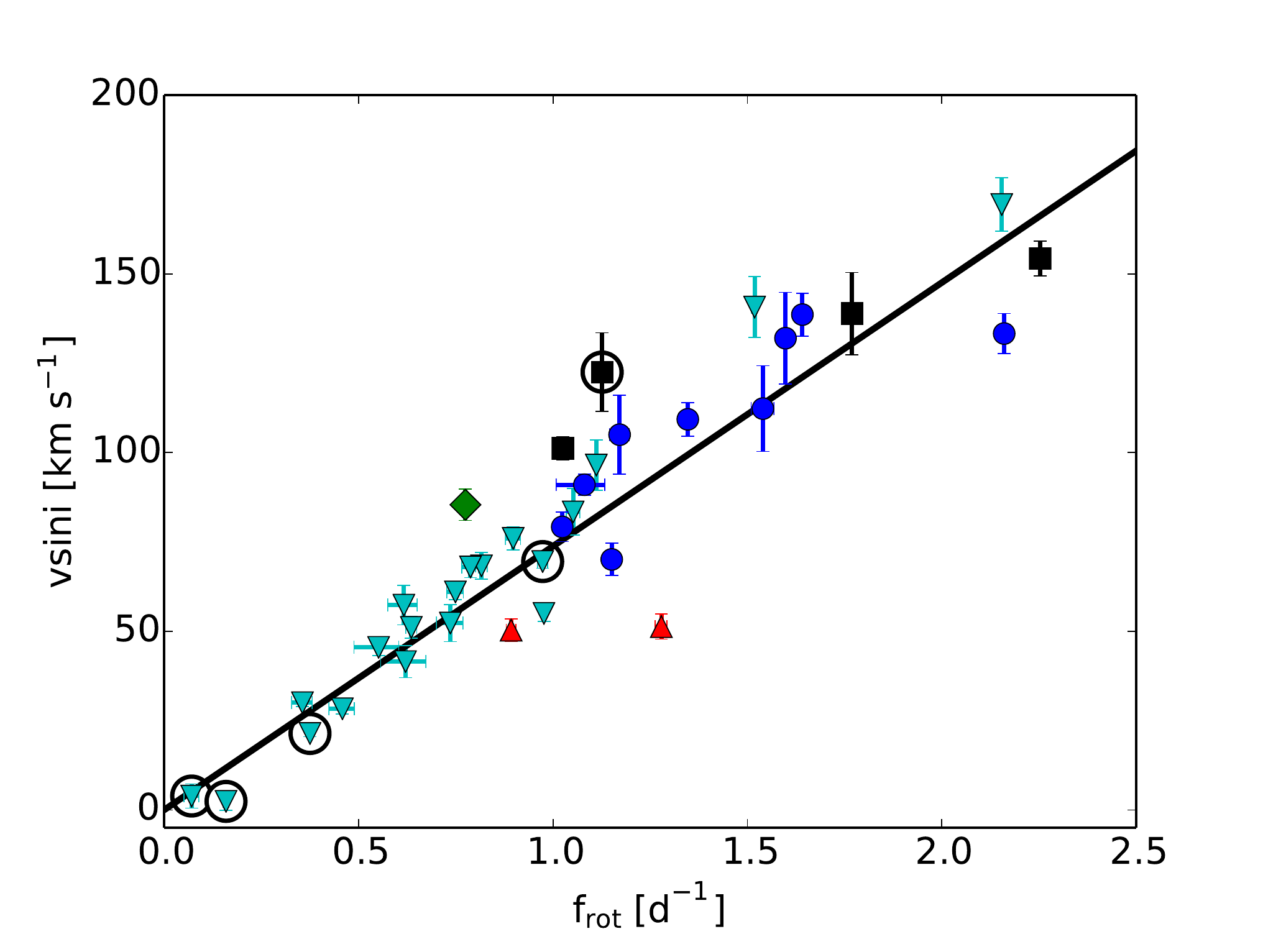}
 \caption{\label{fig:vsini_frot} The correlation between the
     spectroscopic $v\sin i$ values and the values of the rotation rate $f_{\rm
       rot}$ which were derived in this study. The black line indicates the
     corresponding linear fit, for which the coefficients are listed in
     Table\,\ref{tab:correll}. The used symbols are the same as in
     Fig.\,\ref{fig:sample_ov}. For all of the stars a series of prograde dipole
     modes was detected. If another series was detected as well, the symbol of
     the corresponding mode identification was used.}
\end{figure}

Most of the correlations presented previously by \citet{VanReeth2015} were
indicative of the strong relation between the observed gravity-mode pulsations and the
stellar rotation. In retrospect, these can now be linked to the identification
of most pulsations as prograde dipole gravity, gravito-inertial or
  retrograde Rossby modes with $|m| = 1$. In particular, those previous results
are echoed in our current work by the detected correlations between $f_{\rm rot}$
and $v\sin i$, and $f_{\rm rot}$ and $f_{\rm dom,inert}$. The strong correlation
between $f_{\rm rot}$ and $v\sin i$ is illustrated in
Fig.\,\ref{fig:vsini_frot}. The previously detected correlations between
  $v\sin i$ and the mean period spacing $\langle\Delta P\rangle$, the mean
  pulsation period $\langle P\rangle$ and the mean slope
  $\langle\frac{\mathrm{d}\Delta P}{\mathrm{d}P}\rangle$ of the observed series
  discussed in \citet{VanReeth2015} are now also reflected in similar
  correlations with $f_{\rm rot}$. We do find a level of scatter in the
  relationship between these parameters and the rotation rate $f_{\rm rot}$, originating
  from the large variety of radial orders of the detected modes, the limited lengths of some of the observed series and from
  non-uniform variations in the period spacing patterns covered by our sample.
  Moreover, we have assumed a constant rotation rate throughout the stars to
  deduce $f_{\rm rot}$, which is simplistic compared to predictions based on
  numerical simulations \citep{Rogers2015}. Allowing for a variety of
  non-uniform interior rotation profiles will likely complicate the
  correlations.

\citet{VanReeth2015} also found a smaller contribution of $T_{\rm eff}$ to the
multivariate correlation with $f_{\rm dom,inert}$ and $v\sin i$. While this
contribution drops when we replace $v\sin i$ with $f_{\rm rot}$, there is a weak
correlation between $T_{\rm eff}$ and $R\sin i$. Indeed, as a star ages, its
temperature $T_{\rm eff}$ drops and its radius increases. A similar weak
correlation was found between $f_{\rm rot}$ and $\log\,g$, indicating that as
the star evolves and its radius increases, both the surface gravity and the
rotation rate decrease. The correlation between $R\sin i$ and $\log\,g$ was not
significant, likely due to the relatively large uncertainties.

In contrast, we did not find correlations between the asymptotic spacing
$\Delta\Pi_l$ and any of the other parameters. The uncertainty margins on the
value of $\Delta\Pi_l$ are likely too large for a proper correlation to be
unravelled. Multivariate correlations were not detected either.

\begin{table*}
 \caption{\label{tab:correll} Results of the linear regression analysis. We list the coefficients of the covariates for the
different correlations as well as their $p$-values (obtained from a $t$-test) and $R^2$ values.}
\centering
\begin{tabular}{llllll}
\hline\hline
Explanatory variable & Dependent variable & Intercept ($\sigma$) & Estimate ($\sigma$) & $p$-value & $R^2$\\
\hline
$f_{\rm rot}$ [$\rm d^{-1}$] & $v\sin i$ [$\rm km\,s^{-1}$] & 0(16) & 74(5) & $< 0.0001$ & 0.859\\
$f_{\rm rot}$ [$\rm d^{-1}$]& $f_{\rm dom,inert}$ [$\rm d^{-1}$]& 0.7(0.3) & 0.90(0.08) & $< 0.0001$ & 0.780\\
$f_{\rm rot}$ [$\rm d^{-1}$]& $\langle P\rangle$ [$\rm d$]& 0.9(0.1) & -0.22(0.06) & $< 0.0001$ & 0.630\\
$f_{\rm rot}$ [$\rm d^{-1}$]& $\langle \Delta P\rangle$ [$\rm d$]& 0.017(0.005) & -0.009(0.002) & $< 0.0001$ & 0.508\\
$f_{\rm rot}$ [$\rm d^{-1}$]& $\langle \frac{\mathrm{d}\Delta P}{\mathrm{d}P}\rangle$ & -0.013(0.006) & -0.013(0.001)  & $< 0.0001$ & 0.479\\
$T_{\rm eff}$ [K]& $R\sin i$ [$R_\odot$] & 573(10) &  -0.08(0.02) & 0.0001 & 0.357\\
$\log\,g$ [dex]& $f_{\rm rot}$ [$\rm d^{-1}$]& -5.8(0.4) & 1.7(0.4) & 0.0002 & 0.332\\
  \hline
\end{tabular} 
\end{table*}

\section{Discussion and conclusions}
\label{sec:conclusions}
In this paper, we have presented methodology to derive the near-core
  interior rotation rate $f_{\rm
  rot}$ from an observed period spacing pattern and to perform mode
identification for the pulsations in the series.  In a first step, we
considered all combinations of $l$- and $m$-values for mode
  identification. For each pair of $(l,m)$we consider the asymptotic spacing
  $\Delta\Pi_l$ and compute the corresponding equidistant model period spacing
pattern as described by \citet{Tassoul1980}. Using the traditional
approximation, the frequencies of the model pattern are subsequently shifted for
an assumed rotation rate $f_{\rm rot}$ and the chosen $l$ and $m$. The
optimal values of $\Delta\Pi_l$, $f_{\rm rot}$, $l$ and $m$ are then
determined by fitting the model pattern to the observed period spacing
series using least-squares optimisation and taking into account that different
values of $\Delta\Pi_l$ are expected for different values of $l$.

In most cases this method is reasonably successful. For slow rotators it may be
difficult to find the correct value for the azimuthal order $m$, though this
problem is solved when we have multiple series with different $l$ and $m$
values. By fitting these series simultaneously, not only do we obtain the mode
identification, but the values for $\Delta\Pi_l$ and $f_{\rm rot}$ are also a
lot more precise than in the case where we do not detect multiplets. In the case
where we are dealing with a moderate to fast rotator, the retrograde modes
were found to be Rossby modes, which arise due to the interaction between the
stellar rotation and toroidal modes.  In this study, we have used the
asymptotic approximation derived by \citet{Townsend2003c} to compute their
eigenvalues $\lambda$ of the Laplace tidal equation. A complete numerical
treatment of these modes is required to exploit them quantitatively. A
complete and detailed analysis of such stars with multiple gravity-mode period
spacings will allow us to study possible differential rotation in $\gamma$\,Dor
stars, ultimately leading to proper observational constraints on rotational
chemical mixing and angular momentum transport mechanisms.

From the ensemble modelling of the gravity-mode period spacings of the stars in
our sample, we found that there is a large range in the stellar rotation
rates. Interestingly, only three out of forty targets were found to be in the
superinertial regime. These three stars, KIC\,8645874, KIC\,9751996 and
KIC\,11754232, are hybrid $\gamma$\,Dor/$\delta$\,Sct stars which exhibit
variability in the frequency range from 5\,$\rm d^{-1}$ to 8\,$\rm d^{-1}$. This
indicates that these stars' low rotation rates are likely linked to their hybrid
character, making them prime targets for further asteroseismological
analysis. The other stars were found to be in the subinertial regime. Their
pulsation frequencies in the corotating frame are typically confined in the
narrow range between 0.15 and 0.75\,$\rm d^{-1}$. This is in agreement with the
theoretical expectation that $\gamma$\,Dor pulsation frequencies in the
corotating frame are on the order of the thermal timescale $\tau_{th}$ at the
bottom of the convective envelope \citep{Bouabid2013}. However, this frequency
range does not agree with the predicted values by \citet{Bouabid2013}. With the
exception of the three stars in the superinertial regime, we find that on
average the observed modes have longer pulsation periods in the corotating frame
than theory predicts. This is also reflected in the high spin parameter values
we derived for many of the stars. The high spin parameters detected
for the retrograde Rossby modes are linked to the low eigenvalues $\lambda$ of
these modes as already found on theoretical grounds by \citet{Townsend2003c}.

The global results for the mode identification are consistent with existing
spectroscopic studies. The majority of the modes were found to be prograde
dipole modes. This is in line with the results obtained by \citet{Townsend2003b}
for heat-driven gravity modes in slowly pulsating B stars. In addition, we found
single high-amplitude modes, as opposed to a series, to be present in several
stars. They are consistent with retrograde Rossby modes with $m=-1$. They
are likely heavily influenced by mode trapping, and as a result contain valuable
information about these stars' internal structure.

We conducted a linear regression analysis on the combined spectroscopic
and photometric parameter values for the sample. The strong correlation
between $v\sin i$ and $f_{\rm rot}$ independently confirmed the reliability of
the obtained rotation rates. We also detected weak correlations between $R\sin i
= v\sin i/f_{\rm rot}$ and $T_{\rm eff}$ and between $\log\,g$ and $f_{\rm
  rot}$. Indeed, as a star with a convective core evolves on the main sequence,
its radius increases, and its temperature and rotation rate decrease.

Despite the limitations of the traditional approximation, the results we
  obtained in this work are consistent and offer the first estimates of the
  interior rotation frequencies for a large sample of $\gamma\,$Dor stars. The large observed spin
  parameter values indicate that the pulsations are constrained in a
  waveguide around the equator \citep{Townsend2003b,Townsend2003c}. This in turn implies that
  the vast majority of the stars should be seen at moderate to high inclination angles, which is
  also what we can indirectly derive from the relation between the observed
  $v\sin i$ and $f_{\rm rot}$ in Fig.\,\ref{fig:vsini_frot}. From the grid of
  theoretical models in Section\,\ref{sec:grid}, we find radii between
  1.3\,$R_{\odot}$ and 3\,$R_{\odot}$. For many stars in our sample, this
  results in inclination angle estimates on the order of or above
  50\,\textdegree. Two of the stars for which lower inclination angle estimates
  were found, KIC\,4846809 and KIC\,9595743, are also the stars for which we
  detected zonal dipole modes. This is consistent with expectations for the
  geometrical cancellation effects of the pulsations.

These ensemble analyses now form an ideal starting point for detailed
asteroseismological modelling of individual targets in the sample.
This, in turn, will allow us to place constraints on
the shape and extent of the convective core overshooting and the
diffusive mixing processes in the radiative near-core regions, and
by extension on the evolution of the convective core itself as it was recently
achieved for a hybrid $\delta\,$Sct — $\gamma\,$Dor binary \citep{SchmidAerts2016} 
and also for a slowly \citep{Moravveji2015} and a moderately \citep{Moravveji2016} 
rotating gravity-mode pulsator of $\sim$ 3.3\,$M_\odot$.

\begin{acknowledgements}
  The research leading to these results was based on funding from the Fund for
  Scientific Research of Flanders (FWO), Belgium, under grant agreement
  G.0B69.13, and on funding from the European Research Council (ERC) under the
  European Union’s Horizon 2020 research and innovation programme (grant
  agreement N 670519: MAMSIE). TVR thanks Ehsan Moravveji for the extensive
  discussions on the use of the MESA and GYRE codes, and Santiago A.\ Triana for
  the enlightening conversations about the influence of rotation on stellar
  pulsations. TVR also thanks Fran\c{c}ois Ligni\`eres and the other
    participants of the Toulouse 2016 SpaceInn workshop on Stellar Rotation for
    the useful discussions. We further thank the anonymous referee for helpful
    remarks that helped us to improve the interpretations and presentation of
    the research. We are grateful to Bill Paxton and Richard Townsend for their 
    valuable work on the stellar evolution code MESA and stellar pulsation code 
    GYRE. We gratefully acknowledge the Th\"uringer Landessternwarte in
  Tautenburg, Germany, for the computation time on their computer
  cluster. Funding for the \emph{Kepler\hyphenation{Kep-ler}} mission is
  provided by NASA's Science Mission Directorate. We thank the whole team for
  the development and operations of the mission. This research made use of the
  SIMBAD database, operated at CDS, Strasbourg, France, and the SAO/NASA
  Astrophysics Data System. This research has made use of the VizieR catalogue
  access tool, CDS, Strasbourg, France.
\end{acknowledgements}

\bibliographystyle{aa} 
\bibliography{Rotation_paper}
\newpage
\begin{appendix}
\section{Simulated period spacing pattern}
 \begin{table}[h]
 \caption{\label{tab:synthP} The pulsation periods (in days; in the inertial reference frame) of the simulated data in Section \ref{subsec:synthdata}, with the used 1$\sigma$ uncertainty margins.}
\centering
\begin{tabular}{llll}
\hline\hline
$P_{\rm inert}$ [$d$] & $\sigma_P$ & $P_{\rm inert}$ [$d$] & $\sigma_P$\\
\hline
 0.73912 & 0.00004   &   0.91909 & 0.00002\\
0.751414 & 0.000008   &   0.92658 & 0.00006\\
0.76322 & 0.00005   &   0.93385 & 0.00004\\
0.774892 & 0.000007   &   0.94088 & 0.00001\\
0.78640 & 0.00003   &   0.94772 & 0.00001\\
0.79748 & 0.00006   &   0.95442 & 0.00007\\
0.80785 & 0.00001   &   0.96100 & 0.00004\\
0.81753 & 0.00005   &   0.96745 & 0.00006\\
0.82691 & 0.00004   &   0.97373 & 0.00006\\
0.83618 & 0.00004   &   0.97984 & 0.00001\\
0.84513 & 0.00006   &   0.98579 & 0.00006\\
0.85371 & 0.00003   &   0.99161 & 0.00008\\
0.86220 & 0.00006   &   0.99732 & 0.00003\\
0.87080 & 0.00003   &   1.00292 & 0.00008\\
0.87937 & 0.00004   &   1.00839 & 0.00005\\
0.88773 & 0.00008   &   1.01372 & 0.00009\\
0.89582 & 0.00003   &   1.01891 & 0.00004\\
0.903689 & 0.000007   &   1.02400 & 0.00008\\
0.91144 & 0.00001   &   1.02898 & 0.00006\\
  \hline
\end{tabular} 
\end{table}

\newpage
\onecolumn
\section{Stellar rotation rates and mode identification}
\begin{longtab}\small
\begin{longtable}{llllllllll}
\caption{The rotation rates $f_{\rm rot}$ and the asymptotic period spacings $\Delta\Pi_l$ which were computed from the observed period spacing patterns, as well as the mode identification and the dominant pulsation frequency for each series. For the latter we also computed the spin parameter $s$, listed in the final column. The pulsation mode patterns which are marked with ``R'' are retrograde modes, which are most likely Rossby modes. The pulsation modes which are marked with ``S'' are single high-amplitude peaks which were present in the frequency spectra, but clearly separate from the detected period spacing patterns. Neither the retrograde pulsation modes nor the single peaks were actually used in the computations. The corresponding rotation rates and asymptotic period spacings were obtained from the prograde series observed for the same star.}
\label{tab:param} 
\\\hline \hline
\multicolumn{1}{c}{KIC} & \multicolumn{1}{c}{$l$} & \multicolumn{1}{c}{$m$} & \multicolumn{2}{c}{$f_{\rm dom,inert}$} & \multicolumn{2}{c}{$f_{\rm rot}$} & \multicolumn{2}{c}{$\Delta \Pi_l$} & \multicolumn{1}{c}{$|s|$}\\[4pt]
 &  &  & \multicolumn{1}{c}{[$\rm d^{-1}$]} & \multicolumn{1}{c}{[$\mu $Hz]} &  \multicolumn{1}{c}{[$\rm d^{-1}$]} & \multicolumn{1}{c}{[$\mu $Hz]} & \multicolumn{1}{c}{[s]} & \multicolumn{1}{c}{[d]} & \\[4pt]
\hline\\ 
2710594 & $1$ & $1$ & $1.35536^{+0.00001}_{-0.00001}$ & $15.68706^{+0.00012}_{-0.00012}$ & $1.02^{+0.02}_{-0.02}$ & $11.8^{+0.2}_{-0.3}$ & $3370^{+ 330}_{- 310}$ & $0.039^{+0.004}_{-0.004}$&  $6.2^{+0.5}_{-0.5}$\\[4pt]
 & R & R & $0.79908^{+0.00002}_{-0.00002}$ & $9.24856^{+0.00022}_{-0.00022}$ & $1.02^{+0.02}_{-0.02}$ & $11.8^{+0.2}_{-0.3}$ & $3370^{+ 330}_{- 310}$ & $0.039^{+0.004}_{-0.004}$ & $9.1^{+0.7}_{-0.6}$\\[4pt]
3448365 & $1$ & $1$ & $1.500157^{+0.000009}_{-0.000009}$ & $17.36293^{+0.00010}_{-0.00010}$ & $1.08^{+0.05}_{-0.07}$ & $12.5^{+0.6}_{-0.8}$ & $3020^{+1160}_{- 970}$ & $0.03^{+0.01}_{-0.01}$&  $5.2^{+0.9}_{-1.2}$\\[4pt]
 & R & R & $0.88877^{+0.00001}_{-0.00001}$ & $10.28675^{+0.00016}_{-0.00016}$ & $1.08^{+0.05}_{-0.07}$ & $12.5^{+0.6}_{-0.8}$ & $3020^{+1160}_{- 970}$ & $0.03^{+0.01}_{-0.01}$ & $11^{+4}_{-3}$\\[4pt]
4846809 & $1$ & $1$ & $1.81324^{+0.00001}_{-0.00001}$ & $20.98662^{+0.00014}_{-0.00014}$ & $1.28^{+0.01}_{-0.02}$ & $14.8^{+0.2}_{-0.2}$ & $2930^{+ 140}_{- 150}$ & $0.034^{+0.002}_{-0.002}$&  $4.8^{+0.2}_{-0.2}$\\[4pt]
& $1$ & $0$ & $1.00410^{+0.00002}_{-0.00002}$ & $11.62156^{+0.00024}_{-0.00024}$ & $1.28^{+0.01}_{-0.02}$ & $14.8^{+0.2}_{-0.2}$ & $2930^{+ 140}_{- 150}$ & $0.034^{+0.002}_{-0.002}$&  $2.55^{+0.03}_{-0.03}$\\[4pt]
5114382 & $1$ & $1$ & $1.47927^{+0.00002}_{-0.00002}$ & $17.12115^{+0.00020}_{-0.00020}$ & $1.15^{+0.02}_{-0.02}$ & $13.3^{+0.2}_{-0.2}$ & $3070^{+ 300}_{- 290}$ & $0.036^{+0.003}_{-0.003}$&  $7.0^{+0.5}_{-0.6}$\\[4pt]
 & R & R & $0.95265^{+0.00002}_{-0.00002}$ & $11.02607^{+0.00018}_{-0.00018}$ & $1.15^{+0.02}_{-0.02}$ & $13.3^{+0.2}_{-0.2}$ & $3070^{+ 300}_{- 290}$ & $0.036^{+0.003}_{-0.003}$ & $11.6^{+1.0}_{-0.9}$\\[4pt]
5522154 & $1$ & $1$ & $3.009858^{+0.000008}_{-0.000008}$ & $34.83632^{+0.00009}_{-0.00009}$ & $2.154^{+0.004}_{-0.004}$ & $24.93^{+0.05}_{-0.05}$ & $3350^{+  40}_{-  30}$ & $0.0388^{+0.0004}_{-0.0004}$&  $5.03^{+0.03}_{-0.03}$\\[4pt]
5708550 & $1$ & $1$ & $1.11550^{+0.00001}_{-0.00001}$ & $12.91091^{+0.00014}_{-0.00014}$ & $0.82^{+0.01}_{-0.02}$ & $9.4^{+0.2}_{-0.2}$ & $3330^{+ 240}_{- 220}$ & $0.039^{+0.003}_{-0.003}$&  $5.5^{+0.4}_{-0.4}$\\[4pt]
5788623 & $1$ & $1$ & $0.77895^{+0.00001}_{-0.00001}$ & $9.01558^{+0.00014}_{-0.00014}$ & $0.40^{+0.04}_{-0.05}$ & $4.6^{+0.5}_{-0.5}$ & $2800^{+480}_{-440}$ & $0.032^{+0.006}_{-0.005}$&  $2.1^{+0.4}_{-0.5}$\\[4pt]
6468146 & $1$ & $1$ & $1.545700^{+0.000005}_{-0.000005}$ & $17.89004^{+0.00005}_{-0.00005}$ & $0.97^{+0.01}_{-0.01}$ & $11.3^{+0.1}_{-0.2}$ & $3000^{+ 110}_{- 110}$ & $0.035^{+0.001}_{-0.001}$& $3.4^{+0.1}_{-0.1}$ \\[4pt]
6468987 & $1$ & $1$ & $1.998989^{+0.000004}_{-0.000004}$ & $23.13644^{+0.00005}_{-0.00005}$ & $1.598^{+0.009}_{-0.008}$ & $18.49^{+0.10}_{-0.09}$ & $3730^{+ 120}_{- 100}$ & $0.043^{+0.001}_{-0.001}$&  $8.0^{+0.2}_{-0.2}$\\[4pt]
 & R & R & $1.387598^{+0.000009}_{-0.000009}$ & $16.06016^{+0.00011}_{-0.00011}$ & $1.598^{+0.009}_{-0.008}$ & $18.49^{+0.10}_{-0.09}$ & $3730^{+ 120}_{- 100}$ & $0.043^{+0.001}_{-0.001}$ & $15.2^{+0.5}_{-0.6}$\\[4pt]
6678174 & $1$ & $1$ & $1.12777^{+0.00002}_{-0.00002}$ & $13.05290^{+0.00020}_{-0.00020}$ & $0.55^{+0.05}_{-0.06}$ & $6.4^{+0.6}_{-0.7}$ & $3370^{+ 710}_{- 620}$ & $0.039^{+0.008}_{-0.007}$&  $1.9^{+0.4}_{-0.4}$\\[4pt]
6935014 & $1$ & $1$ & $1.20670^{+0.00001}_{-0.00001}$ & $13.96648^{+0.00014}_{-0.00014}$ & $0.79^{+0.02}_{-0.02}$ & $9.1^{+0.2}_{-0.3}$ & $3180^{+ 310}_{- 300}$ & $0.037^{+0.004}_{-0.003}$&  $3.8^{+0.3}_{-0.3}$\\[4pt]
6953103 & $1$ & $1$ & $1.287597^{+0.000008}_{-0.000008}$ & $14.90274^{+0.00009}_{-0.00009}$ & $0.74^{+0.03}_{-0.04}$ & $8.5^{+0.4}_{-0.4}$ & $3560^{+ 490}_{- 440}$ & $0.041^{+0.006}_{-0.005}$&  $2.7^{+0.3}_{-0.3}$\\[4pt]
7023122 & $1$ & $1$ & $1.876108^{+0.000003}_{-0.000003}$ & $21.71421^{+0.00003}_{-0.00003}$ & $0.977^{+0.005}_{-0.005}$ & $11.30^{+0.06}_{-0.06}$ & $3380^{+  40}_{-  30}$ & $0.0391^{+0.0004}_{-0.0004}$&  $2.17^{+0.03}_{-0.03}$\\[4pt]
7365537 & $1$ & $1$ & $2.925633^{+0.000004}_{-0.000004}$ & $33.86150^{+0.00004}_{-0.00004}$ & $2.253^{+0.003}_{-0.003}$ & $26.07^{+0.03}_{-0.03}$ & $3340^{+  30}_{-  30}$ & $0.0387^{+0.0003}_{-0.0003}$&  $6.70^{+0.04}_{-0.04}$\\[4pt]
 & S & S & $1.981016^{+0.000006}_{-0.000006}$ & $22.92843^{+0.00006}_{-0.00006}$ & $2.253^{+0.003}_{-0.003}$ & $26.07^{+0.03}_{-0.03}$ & $3340^{+  30}_{-  30}$ & $0.0387^{+0.0003}_{-0.0003}$&  $16.6^{+0.2}_{-0.2}$\\[4pt]
7380501 & $1$ & $1$ & $0.96329^{+0.00001}_{-0.00001}$ & $11.14922^{+0.00012}_{-0.00012}$ & $0.64^{+0.01}_{-0.01}$ & $7.4^{+0.2}_{-0.2}$ & $2860^{+ 180}_{- 170}$ & $0.033^{+0.002}_{-0.002}$&  $3.9^{+0.2}_{-0.3}$\\[4pt]
7434470 & $1$ & $1$ & $2.542409^{+0.000006}_{-0.000006}$ & $29.42603^{+0.00007}_{-0.00007}$ & $1.769^{+0.005}_{-0.005}$ & $20.47^{+0.06}_{-0.06}$ & $3020^{+  50}_{-  50}$ & $0.0349^{+0.0006}_{-0.0005}$&  $4.57^{+0.04}_{-0.04}$\\[4pt]
 & S & S & $1.698729^{+0.000001}_{-0.000001}$ & $19.66122^{+0.00002}_{-0.00002}$ & $1.769^{+0.005}_{-0.005}$ & $20.47^{+0.06}_{-0.06}$ & $3020^{+  50}_{-  50}$ & $0.0349^{+0.0006}_{-0.0005}$&  $50^{+3}_{-4}$\\[4pt]
7583663 & $1$ & $1$ & $1.47213^{+0.00001}_{-0.00001}$ & $17.03853^{+0.00017}_{-0.00017}$ & $1.17^{+0.02}_{-0.03}$ & $13.6^{+0.3}_{-0.3}$ & $3120^{+ 390}_{- 360}$ & $0.036^{+0.005}_{-0.004}$&  $7.8^{+0.8}_{-0.9}$\\[4pt]
 & R & R & $1.044741^{+0.000008}_{-0.000008}$ & $12.09190^{+0.00009}_{-0.00009}$ & $1.17^{+0.02}_{-0.03}$ & $13.6^{+0.3}_{-0.3}$ & $3120^{+ 390}_{- 360}$ & $0.036^{+0.005}_{-0.004}$ & $19^{+3}_{-3}$\\[4pt]
7746984 & $1$ & $1$ & $2.00305^{+0.00002}_{-0.00002}$ & $23.18341^{+0.00019}_{-0.00019}$ & $1.49^{+0.02}_{-0.02}$ & $17.2^{+0.2}_{-0.2}$ & $3130^{+ 250}_{- 230}$ & $0.036^{+0.003}_{-0.003}$&  $5.8^{+0.3}_{-0.3}$\\[4pt]
 & S & S & $1.35180^{+0.00001}_{-0.00001}$ & $15.64582^{+0.00014}_{-0.00014}$ & $1.49^{+0.02}_{-0.02}$ & $17.2^{+0.2}_{-0.2}$ & $3130^{+ 250}_{- 230}$ & $0.036^{+0.003}_{-0.003}$&  $22^{+3}_{-2}$\\[4pt]
7939065 & $1$ & $1$ & $1.728171^{+0.000007}_{-0.000007}$ & $20.00198^{+0.00008}_{-0.00008}$ & $1.111^{+0.006}_{-0.006}$ & $12.86^{+0.07}_{-0.07}$ & $3000^{+  40}_{-  40}$ & $0.0347^{+0.0005}_{-0.0005}$&  $3.60^{+0.05}_{-0.05}$\\[4pt]
8364249 & $1$ & $1$ & $1.869376^{+0.000005}_{-0.000005}$ & $21.63629^{+0.00005}_{-0.00005}$ & $1.519^{+0.007}_{-0.008}$ & $17.58^{+0.08}_{-0.09}$ & $3090^{+ 110}_{- 110}$ & $0.036^{+0.001}_{-0.001}$&  $8.7^{+0.2}_{-0.2}$\\[4pt]
8375138 & $1$ & $1$ & $2.077771^{+0.000007}_{-0.000007}$ & $24.04828^{+0.00008}_{-0.00008}$ & $1.64^{+0.01}_{-0.01}$ & $19.0^{+0.1}_{-0.1}$ & $2930^{+ 150}_{- 150}$ & $0.034^{+0.002}_{-0.002}$&  $7.5^{+0.3}_{-0.3}$\\[4pt]
 & R & R & $1.407115^{+0.000010}_{-0.000010}$ & $16.28606^{+0.00011}_{-0.00011}$ & $1.64^{+0.01}_{-0.01}$ & $19.0^{+0.1}_{-0.1}$ & $2930^{+ 150}_{- 150}$ & $0.034^{+0.002}_{-0.002}$ & $14.0^{+0.7}_{-0.6}$\\[4pt]
8645874 & $1$ & $1$ & $1.847014^{+0.000004}_{-0.000004}$ & $21.37747^{+0.00005}_{-0.00005}$ & $0.375^{+0.002}_{-0.002}$ & $4.34^{+0.02}_{-0.02}$ & $3200^{+ 10}_{-10}$ & $0.0371^{+0.0001}_{-0.0001}$&  $0.510^{+0.003}_{-0.003}$\\[4pt]
8836473 & $1$ & $1$ & $1.88341^{+0.00001}_{-0.00001}$ & $21.79871^{+0.00012}_{-0.00012}$ & $1.13^{+0.01}_{-0.01}$ & $13.0^{+0.2}_{-0.2}$ & $2900^{+ 100}_{- 100}$ & $0.034^{+0.001}_{-0.001}$&  $2.98^{+0.09}_{-0.09}$\\[4pt]
 & S & S & $0.52525^{+0.00001}_{-0.00001}$ & $6.07932^{+0.00015}_{-0.00015}$ & $1.13^{+0.01}_{-0.01}$ & $13.0^{+0.2}_{-0.2}$ & $2900^{+ 100}_{- 100}$ & $0.034^{+0.001}_{-0.001}$&  $3.75^{+0.04}_{-0.04}$\\[4pt]
9210943 & $1$ & $1$ & $2.190853^{+0.000004}_{-0.000004}$ & $25.35710^{+0.00005}_{-0.00005}$ & $1.728^{+0.007}_{-0.010}$ & $19.99^{+0.09}_{-0.11}$ & $3340^{+  80}_{- 100}$ & $0.0386^{+0.0010}_{-0.0012}$&  $7.5^{+0.2}_{-0.2}$\\[4pt]
 & R & R & $1.443566^{+0.000009}_{-0.000009}$ & $16.70794^{+0.00011}_{-0.00011}$ & $1.728^{+0.007}_{-0.010}$ & $19.99^{+0.09}_{-0.11}$ & $3340^{+  80}_{- 100}$ & $0.0386^{+0.0010}_{-0.0012}$ & $12.2^{+0.3}_{-0.3}$\\[4pt]
9480469 & $1$ & $1$ & $1.994822^{+0.000010}_{-0.000010}$ & $23.08821^{+0.00011}_{-0.00011}$ & $1.54^{+0.03}_{-0.03}$ & $17.8^{+0.3}_{-0.3}$ & $2990^{+ 410}_{- 360}$ & $0.035^{+0.005}_{-0.004}$&  $6.8^{+0.5}_{-0.6}$\\[4pt]
 & R & R & $1.32598^{+0.00001}_{-0.00001}$ & $15.34704^{+0.00012}_{-0.00012}$ & $1.54^{+0.03}_{-0.03}$ & $17.8^{+0.3}_{-0.3}$ & $2990^{+ 410}_{- 360}$ & $0.035^{+0.005}_{-0.004}$ & $14^{+2}_{-2}$\\[4pt]
9595743 & $1$ & $1$ & $1.43459^{+0.00002}_{-0.00002}$ & $16.6040^{+0.0002}_{-0.0002}$ & $0.89^{+0.01}_{-0.01}$ & $10.3^{+0.1}_{-0.1}$ & $3050^{+ 110}_{- 110}$ & $0.035^{+0.001}_{-0.001}$&  $3.3^{+0.1}_{-0.1}$\\[4pt]
& $1$ & $0$ & $1.16055^{+0.00002}_{-0.00002}$ & $13.43228^{+0.00019}_{-0.00019}$ & $0.89^{+0.01}_{-0.01}$ & $10.3^{+0.1}_{-0.1}$ & $3050^{+ 110}_{- 110}$ & $0.035^{+0.001}_{-0.001}$&  $1.54^{+0.02}_{-0.02}$\\[4pt]
9751996 & $1$ & $1$ & $1.35387^{+0.00002}_{-0.00002}$ & $15.6698^{+0.0002}_{-0.0002}$ & $0.0696^{+0.0008}_{-0.0008}$ & $0.805^{+0.010}_{-0.010}$ & $3086^{+   6}_{-   6}$ & $0.03572^{+0.00007}_{-0.00007}$&  $0.11^{+0.03}_{-0.03}$\\[4pt]
 & $1$ & $0$ & $1.02805^{+0.00002}_{-0.00002}$ & $11.8987^{+0.0002}_{-0.0002}$ & $0.0696^{+0.0008}_{-0.0008}$ & $0.805^{+0.010}_{-0.010}$ & $3086^{+   6}_{-   6}$ & $0.03572^{+0.00007}_{-0.00007}$&  $0.14^{+0.03}_{-0.03}$\\[4pt]
 & $1$ & $-1$ & $1.28331^{+0.00002}_{-0.00002}$ & $14.8531^{+0.0002}_{-0.0002}$ & $0.0696^{+0.0008}_{-0.0008}$ & $0.805^{+0.010}_{-0.010}$ & $3086^{+   6}_{-   6}$ & $0.03572^{+0.00007}_{-0.00007}$&  $0.11^{+0.03}_{-0.03}$\\[4pt]
10256787 & $1$ & $1$ & $1.077489^{+0.000008}_{-0.000008}$ & $12.47094^{+0.00009}_{-0.00009}$ & $0.59^{+0.04}_{-0.04}$ & $6.9^{+0.4}_{-0.5}$ & $2730^{+ 560}_{- 490}$ & $0.032^{+0.006}_{-0.006}$&  $2.5^{+0.3}_{-0.4}$\\[4pt]
10467146 & $1$ & $1$ & $0.954976^{+0.000009}_{-0.000009}$ & $11.05297^{+0.00011}_{-0.00011}$ & $0.62^{+0.03}_{-0.04}$ & $7.1^{+0.4}_{-0.5}$ & $2940^{+ 600}_{- 540}$ & $0.034^{+0.007}_{-0.006}$&  $3.6^{+0.6}_{-0.7}$\\[4pt]
11080103 & $1$ & $1$ & $1.241393^{+0.000005}_{-0.000005}$ & $14.36797^{+0.00006}_{-0.00006}$ & $0.62^{+0.05}_{-0.06}$ & $7.2^{+0.6}_{-0.7}$ & $3360^{+ 880}_{- 730}$ & $0.039^{+0.010}_{-0.008}$&  $2.0^{+0.3}_{-0.4}$\\[4pt]
11099031 & $1$ & $1$ & $1.61508^{+0.00001}_{-0.00001}$ & $18.69302^{+0.00014}_{-0.00014}$ & $1.025^{+0.009}_{-0.009}$ & $11.87^{+0.10}_{-0.10}$ & $3560^{+ 100}_{- 110}$ & $0.041^{+0.001}_{-0.001}$&  $3.48^{+0.08}_{-0.08}$\\[4pt]
 & S & S & $0.916646^{+0.000009}_{-0.000009}$ & $10.60933^{+0.00010}_{-0.00010}$ & $1.025^{+0.009}_{-0.009}$ & $11.87^{+0.10}_{-0.10}$ & $3560^{+ 100}_{- 110}$ & $0.041^{+0.001}_{-0.001}$&  $19^{+1}_{-1}$\\[4pt]
11294808 & $1$ & $1$ & $1.16617^{+0.00001}_{-0.00001}$ & $13.49738^{+0.00014}_{-0.00014}$ & $0.77^{+0.02}_{-0.02}$ & $9.0^{+0.2}_{-0.3}$ & $2770^{+ 350}_{- 320}$ & $0.032^{+0.004}_{-0.004}$&  $4.0^{+0.3}_{-0.4}$\\[4pt]
 & $2$ & $2$ & $2.2247^{+0.0007}_{-0.0007}$ & $25.749^{+0.008}_{-0.008}$ & $0.77^{+0.02}_{-0.02}$ & $9.0^{+0.2}_{-0.3}$ & $1600^{+ 200}_{--190}$ & $0.019^{+0.002}_{-0.002}$&  $2.3^{+0.2}_{-0.2}$\\[4pt]
11456474 & $1$ & $1$ & $1.471468^{+0.000006}_{-0.000006}$ & $17.03088^{+0.00007}_{-0.00007}$ & $1.05^{+0.02}_{-0.02}$ & $12.2^{+0.2}_{-0.2}$ & $2810^{+ 250}_{- 240}$ & $0.033^{+0.003}_{-0.003}$&  $5.0^{+0.3}_{-0.3}$\\[4pt]
11721304 & $1$ & $1$ & $0.92287^{+0.00002}_{-0.00002}$ & $10.68141^{+0.00020}_{-0.00020}$ & $0.46^{+0.03}_{-0.03}$ & $5.3^{+0.4}_{-0.4}$ & $3080^{+ 430}_{- 400}$ & $0.036^{+0.005}_{-0.005}$&  $2.0^{+0.3}_{-0.3}$\\[4pt]
11754232 & $1$ & $1$ & $1.10338^{+0.00001}_{-0.00001}$ & $12.77065^{+0.00017}_{-0.00017}$ & $0.159^{+0.007}_{-0.007}$ & $1.84^{+0.08}_{-0.09}$ & $3130^{+  20}_{-  20}$ & $0.0362^{+0.0002}_{-0.0003}$&  $0.34^{+0.02}_{-0.02}$\\[4pt]
11826272 & $1$ & $1$ & $0.83370^{+0.00001}_{-0.00001}$ & $9.64927^{+0.00013}_{-0.00013}$ & $0.36^{+0.02}_{-0.03}$ & $4.1^{+0.3}_{-0.3}$ & $2950^{+ 290}_{- 280}$ & $0.034^{+0.003}_{-0.003}$&  $1.5^{+0.2}_{-0.2}$\\[4pt]
11907454 & $1$ & $1$ & $1.77890^{+0.00001}_{-0.00001}$ & $20.58915^{+0.00016}_{-0.00016}$ & $1.35^{+0.02}_{-0.02}$ & $15.6^{+0.2}_{-0.3}$ & $3050^{+ 290}_{- 280}$ & $0.035^{+0.003}_{-0.003}$&  $6.2^{+0.4}_{-0.4}$\\[4pt]
 & R & R & $1.187153^{+0.000008}_{-0.000008}$ & $13.74020^{+0.00009}_{-0.00009}$ & $1.35^{+0.02}_{-0.02}$ & $15.6^{+0.2}_{-0.3}$ & $3050^{+ 290}_{- 280}$ & $0.035^{+0.003}_{-0.003}$&  $17^{+2}_{-2}$\\[4pt]
11917550 & $1$ & $1$ & $1.287680^{+0.000008}_{-0.000008}$ & $14.90371^{+0.00010}_{-0.00010}$ & $0.90^{+0.02}_{-0.02}$ & $10.4^{+0.2}_{-0.2}$ & $2900^{+ 220}_{- 210}$ & $0.034^{+0.003}_{-0.002}$&  $4.6^{+0.3}_{-0.3}$\\[4pt]
11920505 & $1$ & $1$ & $1.198844^{+0.000009}_{-0.000009}$ & $13.87550^{+0.00010}_{-0.00010}$ & $0.75^{+0.02}_{-0.02}$ & $8.7^{+0.2}_{-0.3}$ & $2980^{+ 260}_{- 250}$ & $0.035^{+0.003}_{-0.003}$&  $3.3^{+0.2}_{-0.3}$\\[4pt]
12066947 & $1$ & $1$ & $2.72379^{+0.00001}_{-0.00001}$ & $31.52539^{+0.00015}_{-0.00015}$ & $2.160^{+0.008}_{-0.008}$ & $25.00^{+0.09}_{-0.10}$ & $2950^{+  70}_{-  70}$ & $0.0342^{+0.0008}_{-0.0008}$&  $7.7^{+0.1}_{-0.1}$\\[4pt]
 & R & R & $1.88748^{+0.00002}_{-0.00002}$ & $21.8459^{+0.0002}_{-0.0002}$ & $2.160^{+0.008}_{-0.008}$ & $25.00^{+0.09}_{-0.10}$ & $2950^{+  70}_{-  70}$ & $0.0342^{+0.0008}_{-0.0008}$&  $15.8^{+0.4}_{-0.4}$\\[4pt]
\hline\vspace{30pt}
\end{longtable}
\end{longtab}

\section{Sample analysis}
\begin{figure*}[h]
 \includegraphics[width=\textwidth]{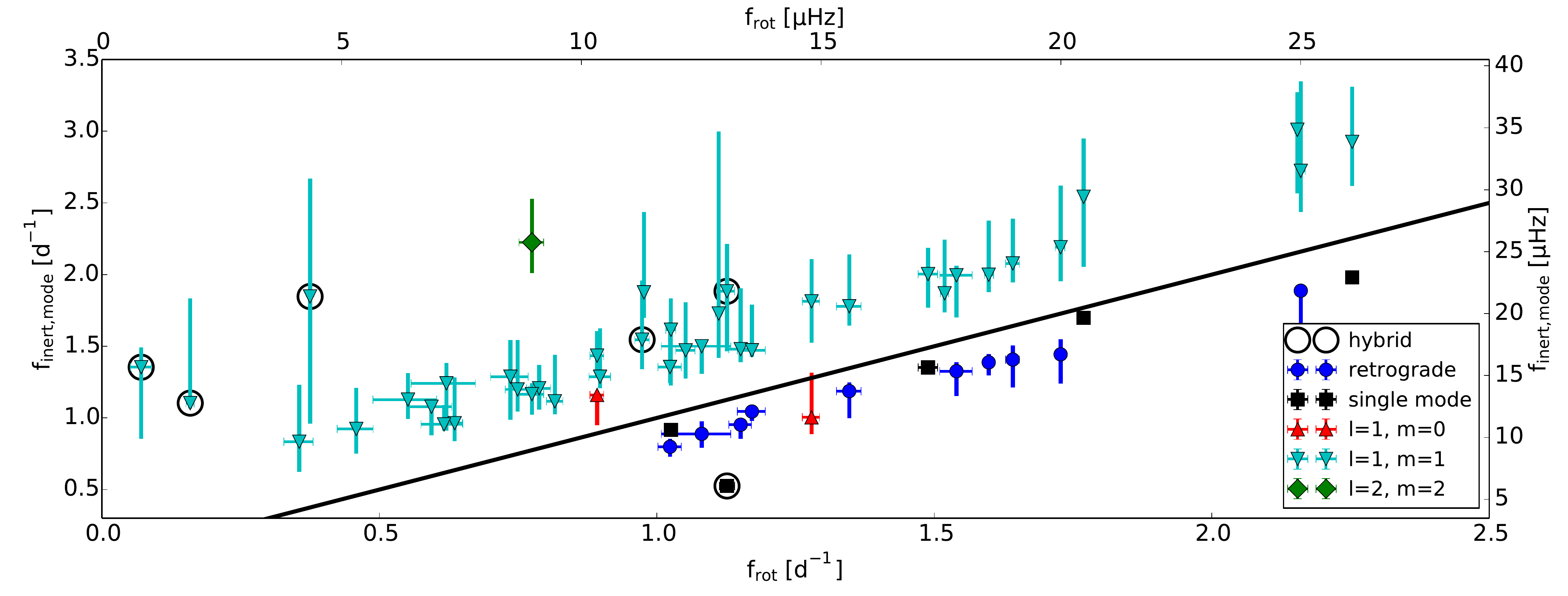}
 \caption{\label{fig:sample_inext} The dominant pulsation frequency $f_{dom,inert}$ in the inertial frame as a function of the rotation frequency $f_{rot}$ for the identified g-mode pulsations of 40 stars in the sample. The thick vertical lines indicate the full extent of the detected spacing series. The full black line indicates where $f_{\rm inert}$ is equal to $f_{\rm rot}$.}
\end{figure*}
\end{appendix}

\end{document}